\documentclass[10pt,aps,pra,twocolumn,amsmath,amssymb,superscriptaddress]{revtex4-1}
\usepackage[english]{babel}
\usepackage{latexsym}
\usepackage{graphics}
\usepackage{epsfig}
\usepackage{color}
\usepackage{bm}
\usepackage{amsmath}
\usepackage{amssymb}
\usepackage{hyperref}

\usepackage[normalem]{ulem}
\usepackage{color}

\newcommand{\ket}[1]{|#1\rangle}

\newcommand{\bb}[1]{\left( #1 \right)}

\newcommand{\be}{\begin{equation}}
\newcommand{\ee}{\end{equation}}
\newcommand{\bea}{\begin{eqnarray}}
\newcommand{\eea}{\end{eqnarray}}

\newcommand{\1}{\uparrow}

\usepackage[T1]{fontenc} 

\begin{document}

\title{Many-body soliton-like states of the bosonic ideal gas}
\author{R.~ O{\l}dziejewski}
\author{W.~G{\'{o}}recki}
\author{K.~ Paw{\l}owski}
\author{K.~Rz\k{a}\.{z}ewski}

\affiliation{Center for Theoretical Physics, Polish Academy of Sciences, Al. Lotnik\'{o}w 32/46, 02-668 Warsaw, Poland}


\begin{abstract}
We study the lowest energy states for fixed total momentum, i.e. yrast states, of $N$ bosons moving on a ring.
As in the paper of A. Syrwid and K. Sacha \cite{syrwid2015}, we compare mean field solitons with the yrast states, being the many-body Lieb-Liniger eigenstates.
We show that even in the limit of vanishing interaction the yrast states possess features typical for solitons, like phase jumps and  density notches. These properties are simply effects of the bosonic symmetrization and are encoded in the Dicke states hidden in the yrast states.
\end{abstract}


\maketitle

\section{Introduction}
It is hard to list all important features, discoveries and applications associated with solitons. These mathematical objects, certain types of solutions of nonlinear integrable differential equations, were found in many areas of Science, ranging from physics to biology and medicine.
\begin{center}
	\begin{figure}[h]
		\includegraphics[width=0.22\textwidth]{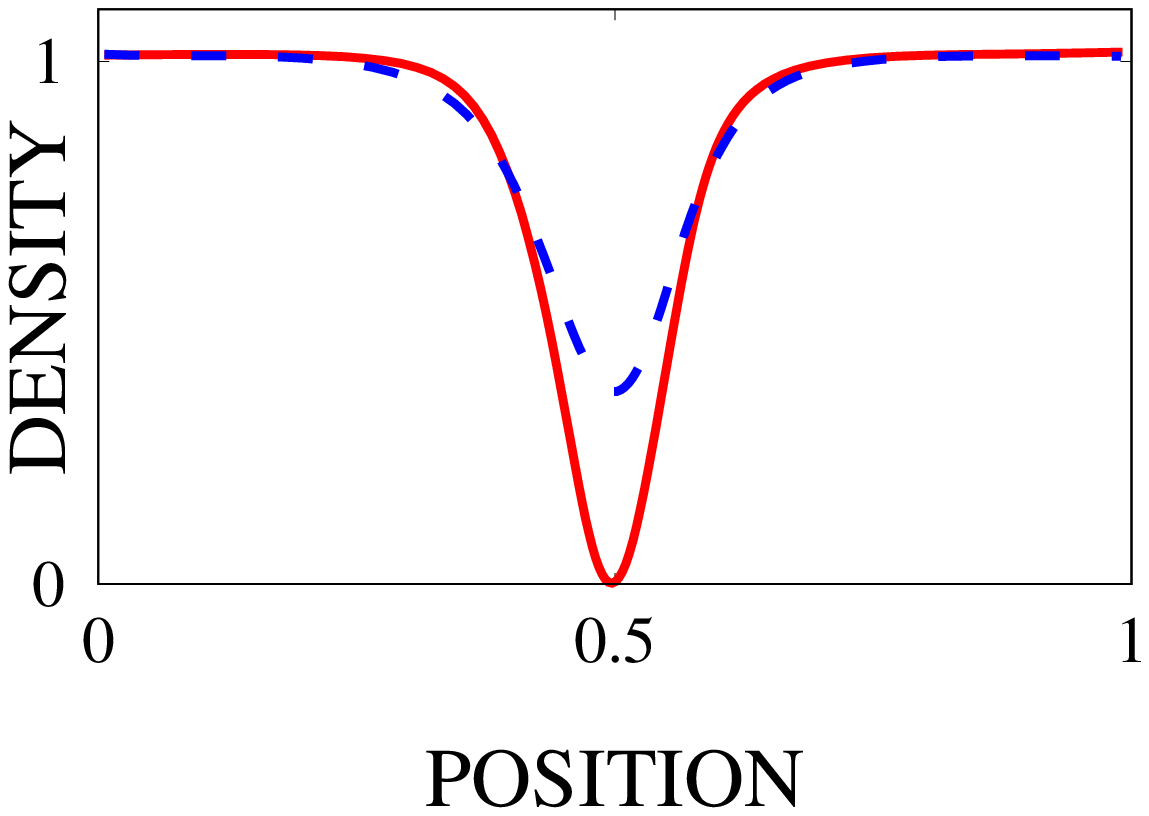}
		\includegraphics[width=0.22\textwidth]{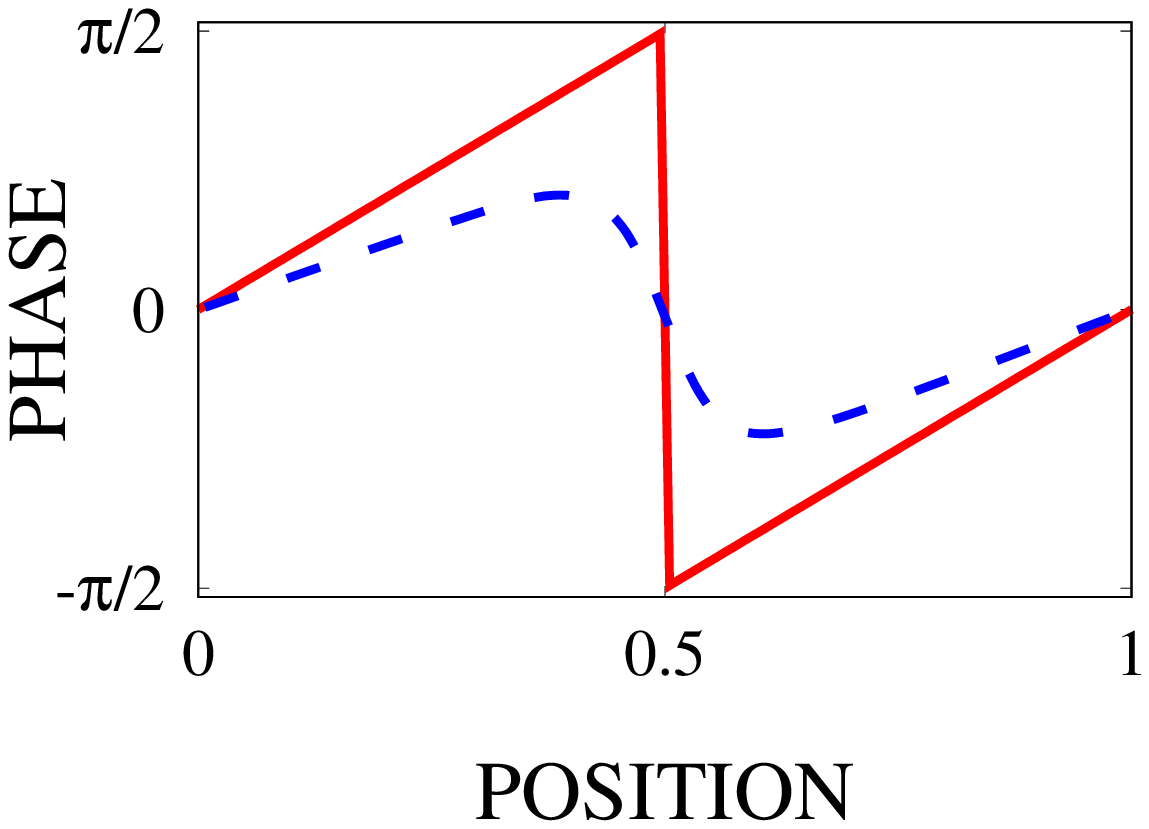}
		
		\caption{(color online) The sketch of density (left) and phase (right) of dark solitons in a box with periodic boundary conditions. The solid red lines correspond to the extreme situation of a black soliton - its density vanishes at the center, whereas the phase has a $\pi$ jump. The blue dashed lines correspond to an example of a gray soliton with the minimal density $0.36$. Position is in the units of the box length $L$. 
			\label{fig:sketch}}
	\end{figure}
\end{center}
There is a number of known equations supporting the solitonic solutions. In physics, very important examples are the Korteweg-de Vries equation \cite{KortewegVries}, Sine-Gordon equation \cite{Bour1862, Frenkel1939} and the non-linear Schr\"odinger equation  \cite{Gross1961, Pitaevskii1961}. Here we will focus on the last case, in the context of $N$ interacting bosons  called also the Gross-Pitaevskii equation (GPE)  \cite{Gross1961, Pitaevskii1961}:
\begin{equation}\label{eq:non-linear-Schrodinger-equation}
i  \frac{\partial \psi_{\rm GPE}}{\partial\,t}  = - \frac{1}{2}\frac{\partial^2 \psi_{\rm GPE}}{\partial\,x^2} + g N\left| \psi_{\rm GPE}\right|^2 \,\psi_{\rm GPE} \,,
\end{equation}
where $g$ is the coupling strength and we set $\hbar=m=1$. This equation has also  proved 
 to be  useful to describe the electric field of light in the non-linear media \cite{KIVSHAR199881}.
In the context of atoms it is the so called mean field (MF) description of the weakly interacting bosons \cite{Frantzeskakis2010}.
 The solitonic solutions of this equation were derived already in the 70s by A. Shabat and V. Zakharov \cite{zakharov71, zakharov73}.
We recall the main finding for the positive coupling strength, $g>0$.  
In this case the spatial density in the soliton has a single characteristic notch. Within the area  of the notch the phase of $\psi_{\rm GPE}$ is quickly changing. In the extreme situation, the density in the middle of the soliton is zero and the phase has a $\pi$ jump.
The width of the soliton is given by the healing length $\xi = 1/\sqrt{g n} $, where $n$ is the average density of the gas.
The properties of dark solitons are illustrated  in Fig. \ref{fig:sketch}.

The description of the weakly interacting bosonic gas in the frame of the Gross-Pitaevskii equation turned out to be very powerful. Predictions based on this equation have been successfully tested experimentally, including a shape of the Bose-Einstein condensate \cite{hau1998}, its energy, normal modes of excitations and many other nonlinear phenomena.
Shortly after cooling atoms down to the Bose-Einstein condensate regime, also the solitons have been generated \cite{Sengstock, Denschlag2000}. In the present days solitons are routinely produced with the phase imprinting method in many laboratories around the world.

On the other hand Eq. \eqref{eq:non-linear-Schrodinger-equation} provides a  simplified description of $N$ interacting cold atoms.
It is only approximation of the more fundamental many-body linear model in which the state of the system is given by the many-body wave-function depending on positions of all particles. In the case of  $N$ short-range interacting bosons moving on a circumference of the circle of length $L$, their Hamiltonian reads 
	\begin{equation}\label{Ham}
	\hat{H}_{\rm LL} = -\frac{1}{2}\sum_{i=1}^N \frac{\partial^2}{{\partial} x_i^2} + g \sum_{1 \leq i < j \leq N} \delta \bb{x_i-x_j},
	\end{equation}
where $x_i$ is the position of the $i$th boson. The naive derivation of the Gross-Pitaevskii equation is based on the Ansatz, in which one assumes that all particles occupy a single orbital $\psi (x_1,\ldots,x_N,\,t) = \prod_{i=1}^N \, \psi_{\rm GPE} (x_i, \,t)$. Minimization of the time-dependent action averaged in this Ansatz gives the equation for the optimal orbital, Eq. \eqref{eq:non-linear-Schrodinger-equation}.

The gas of atoms described by the Hamiltonian  \eqref{Ham} and under assumption of the periodic boundary condition is called the Lieb-Liniger model \cite{Lieb1963, LiebLiniger1963}. 
This problem has known exact solutions for the eigenstates \cite{LiebLiniger1963,Lieb1963}.
The peculiar thing is, that trying to classify the  excitations in a reasonable way, the Author of \cite{Lieb1963} has found two  types of  elementary excitations. One branch of the elementary excitations was immediately identified with the Bogoliubov excitations. After many years it turned out that the second type of elementary excitations have the same energy - velocity relation as the solitons known from the nonlinear Schr\"odinger equation \cite{Kulish1976}. Further studies \cite{KanamotoCarr2008, KanamotoCarr2010,syrwid2015,zaremba2013, Fialko2012} showed explicitly the correspondence between the type II many-body elementary excitations in the Lieb-Liniger model and the solutions of Eq. \eqref{eq:non-linear-Schrodinger-equation}.
In this paper we follow the ideas of \cite{syrwid2015,syrwid2016}. 
Our goal is to better understand the structure of the type II excitations and the corresponding solutions of the mean field model.
We will demonstrate the role of bosonic statistics in the emergence of the solitonic properties, even in the case without interaction.

The paper is organized as follows. In Sec. \ref{sec:Weakly-interacting case}  we follow the procedure described in the paper \cite{syrwid2015} to extract the mean-field solitons out of the many-body solutions of the Lieb-Liniger Hamiltonian \eqref{Ham}. In particular we find that the structure of the corresponding many-body eigenstates is very simple and close to the case without interaction. This is clarified  in the main part of this paper,  in Sec. \ref{sec:no-interacting-limit}, in which we show that the 
multiparticle configurations with the solitonic properties may be found already among the many-body eigenstates of the Hamiltonian \eqref{Ham} even without interaction, i.e. $g=0$.
We show that the dark, gray and multiple solitons-like states arise already in the non-interacting case and then study their motion
\eqref{subSec:multi-solitons}, extracted from the many-body eigenstates, i.e. time-independent states.

In Sec. \ref{sec:validity-range} we discuss to what extent the conclusions derived for the non-interacting case may be also applied to the interacting gas.

\section{Weakly-interacting gas\label{sec:Weakly-interacting case}}
The purpose of this section is to recall the correspondence between the mean-field solitons and the yrast states of the Lieb-Liniger Hamiltonian \eqref{Ham}, as it has been done in \cite{syrwid2015}.

More than half of the century ago, the Lieb-Liniger Hamiltonian has been solved exactly with the help of the Bethe Ansatz \cite{Lieb1963}. This Ansatz is constructed from plane waves with $N$ parameters, called quasi-momenta, satisfying a set of the transcendental Bethe equations \cite{Lieb1963}. 
The elementary excitations are such many-body eigenstates that differ from the ground state by a single quasi-momentum. 
Depending on the choice of this quasi-momentum the elementary excitations are divided into two groups: the Bogoliubov branch and the type II solitonic branch.

As the system is translationally invariant, the total momentum $\hat{K} = \frac{L}{2\pi i}\sum_{i=1}^N \frac{\partial}{\partial\,x_i}$ commutes with the Hamiltonian. Hence all eigenstates may be numbered by the value of their total momentum $K$. Note that we express the total momentum in units $(2\pi/L)$ so it is an integer.
Our subject of interest are the lowest energy eigenstatetes with a given total momentum, so called yrast states \cite{HAMAMOTO199065, Mottelson1999}. Here we will consider only the contact interaction for which the yrast states coincide with the type II solitonic  elementary excitations (as shown in \cite{NaszRoton} it does not need to be the case for dipolar interactions).

How to extract properties of a single-body wave-function from the many-body eigenstates? 
The naive approach would be to reduce the many-body density matrix by tracing out $N-1$ atoms. This approach would fail -- all eigenstates would be projected to exactly the same single-body uniform density, as a result of the translational invariance.
The Authors of the paper \cite{syrwid2015} have shown another procedure, in the spirit of \cite{Javanainen1996}, which reveals the spatial structures hidden in the eigenstates. One obtains a conditional single-body wave-function by means of drawing remaining $N-1$ particle-positions. 
The position of the first particle $\bar{x}_1$ is drawn from the uniform distribution, $P(x_1)=1/L$. 
Then the position of the second one $\bar{x}_2$ is drawn from the conditional distribution, obtained by setting the first argument of the many-body wave-function as the parameter with the value $\bar{x}_1$ and tracing out the particles $x_3, \, x_4,\,\ldots, x_N$, i.e.  from the distribution $P(x_2) \propto \int |\psi\bb{\bar{x}_1, \, x_2,\,\ldots, x_N}|^2\,\mbox{d}\,x_3\,\mbox{d}\,x_4\ldots\,\mbox{d}\,x_N $. The procedure is repeated until the conditional single-particle  wave-function is reached:
	\begin{equation}
	\psi_{\rm con}^{\bar{x}_1,\, \bar{x}_2,\,\ldots, \bar{x}_{N-1}} (x_N) \propto \psi (\bar{x}_1,\, \bar{x}_2,\,\ldots, \bar{x}_{N-1}, x_N )
	\label{eq:conditional-wave-function}
	\end{equation}
Although the solutions of the Lieb-Liniger model \cite{Lieb1963} are known, it is much more efficient to solve the model numerically. 
We perform calculations in the Fock basis $\ket{n_{-k_{\rm max}},\ldots,n_{k}, \ldots, n_{+k_{\rm max}} }$, with $n_k$ atoms occupying the orbital $e^{i 2 \pi k x/L}$ with an integer  $k$.
 We use the cut-off for maximal momentum $k_{\rm max}$ sufficiently high to ensure convergence. 
The lowest-energy state in the subspace with the total momentum $K$ is found with the discretized form of the imaginary time evolution. To this end, we act repeatedly with the operator $(C-\hat{H})$ on a random state $\ket{\psi_{\rm random}}$, where $C$ is any  constant larger than the maximal eigenvalue of $\hat{H}$ \footnote{As we restrict the states by the maximal cut-off hence $\hat{H}$ is bounded.}
and $\ket{\psi_{\rm random}}$ is any state constructed from the Fock states with the total momentum $K$. In the limit of many repetitions of this operation one obtains the lowest energy state, i.e. $\lim_{n\to \infty} (C-\hat{H})^n \ket{\psi_{\rm random}}$ converges to the yrast state (up to a normalization factor).
\begin{center}
	\begin{figure}[h]
	  \includegraphics[width=0.22\textwidth]{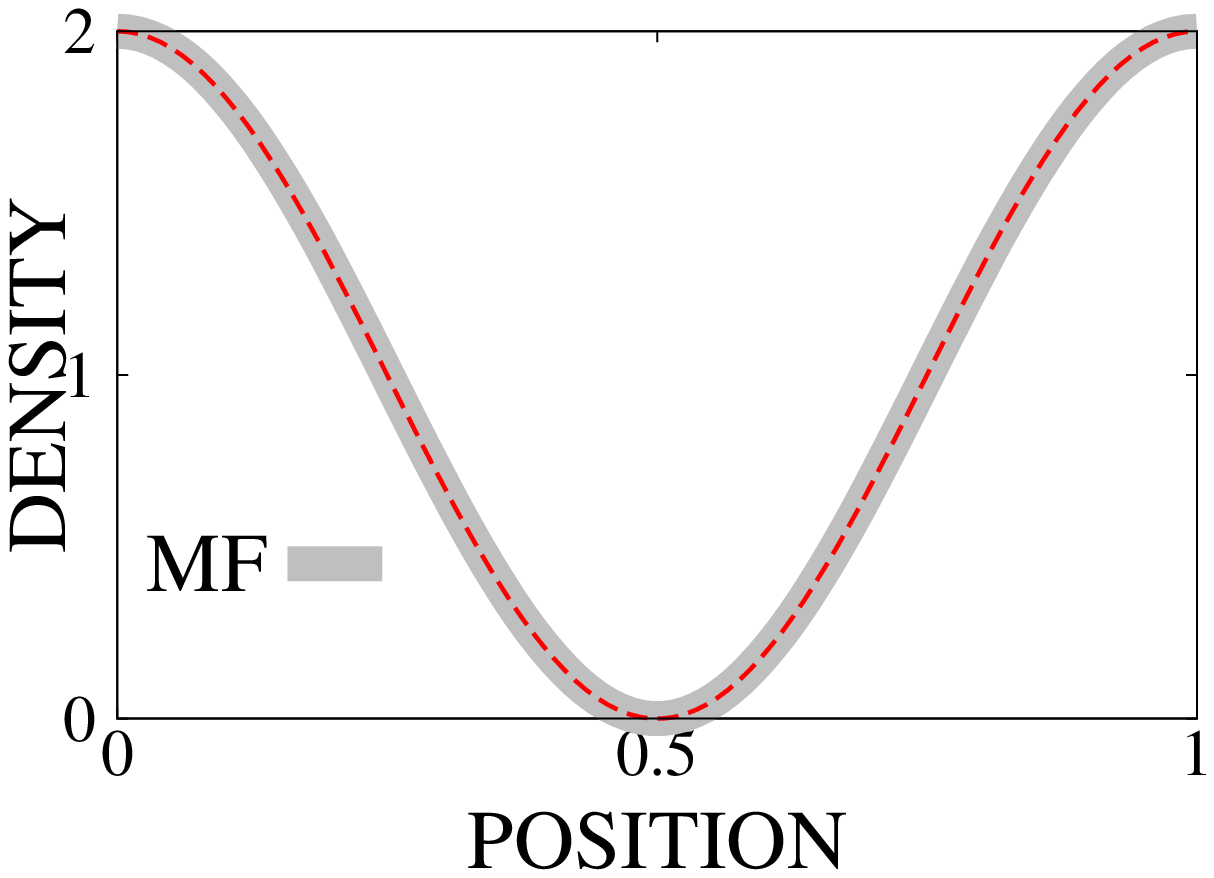}
	  \includegraphics[width=0.22\textwidth]{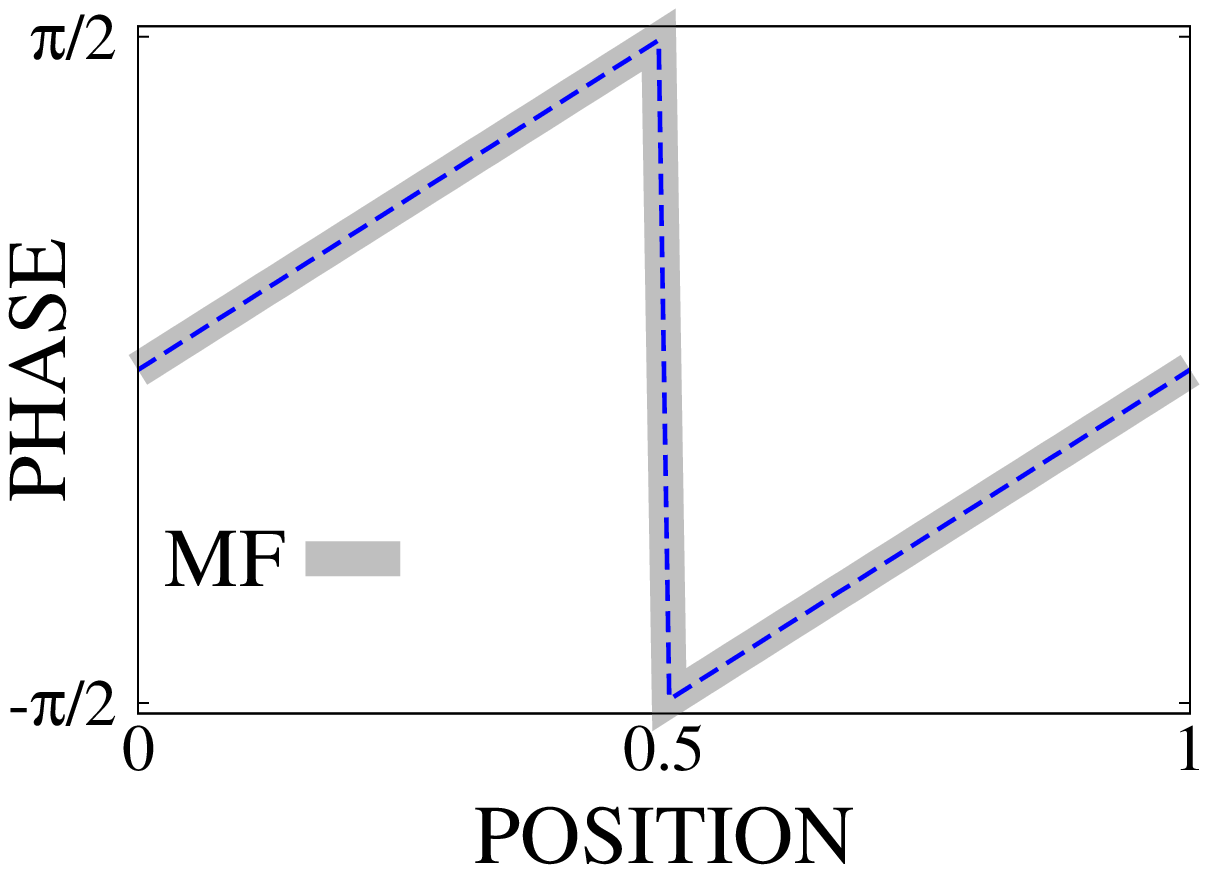}
	  	  
		\caption{(color online) The probability density (left) and the phase (right) of the single-particle conditional wave-function obtained from the many-body yrast state with the total momentum $K=N/2$ by drawing $N-1$ positions, as described in the main text. The corresponding properties of the mean-field solutions are presented with the gray thick line.
		Total number of atoms $N=8$, the interaction strength $g=0.08$. Position is in the units of the box length $L$.
		\label{fig:sol}}
	\end{figure}
\end{center}
In  Fig. \ref{fig:sol} we show an example of the probability distribution of the last particle $P(x_N) \propto |\psi_{\rm con}^{\bar{x}_1,\, \bar{x}_2,\,\ldots, \bar{x}_{N-1}} (x_N)|^2$ and the phase of the wave-function \eqref{eq:conditional-wave-function}, compared with the corresponding quantities of the black soliton -- obtained  from Eq. \eqref{eq:non-linear-Schrodinger-equation}. 
The solutions of the  non-linear Schr\"odinger equation were found with the help of the paper \cite{kanamoto2009} (see also \cite{Carr2000, zaremba2013}). These solutions are given in terms of the elliptic Jacobi functions. We plot the mean field solutions with the average momentum $\frac{L}{2 \pi i}\int \psi_{\rm GPE}^* \frac{\partial}{\partial\,x}\psi_{\rm GPE}\,{\rm d} x$ equal to the total momentum of the yrast state per particle $K/N$.
In the Fig. \ref{fig:sol} we only repeat the result of \cite{syrwid2015}, the one for the weakest interaction.

Our computation performed in the Fock basis gives us immediately access to the structure of the state. 
It turns out that the many-body yrast state from which we obtained the black soliton is dominated by the single Fock state $\ket{n_0 = \frac{N}{2}, n_{1} = \frac{N}{2}}$ with atoms equally distributed between the orbitals with momenta $k=0$ and $k=1$.
The fidelity of this single Fock state and the total state exceeds $99.5$\% \footnote{For this reason Fig 3 was obtained for the ideal gas as it would not be different to weakly interacting gas with ${gN = 0.64}$}. 
The dominant role of the  Fock state for weakly interacting gas is already established in the literature \cite{Fialko2012}.
In the following sections we will discuss how this Fock state is related to the mean-field solitons.  Interesting insight can be reached already in the limit of the ideal gas.


\section{The non-interacting limit\label{sec:no-interacting-limit}}
\subsection{Two branches of excitations \label{subSec:two-branches}}

	\begin{figure}[h]
		\includegraphics[width=0.5\textwidth]{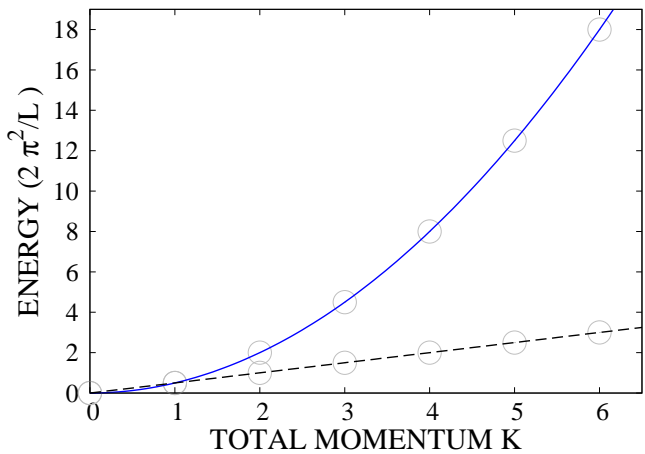}

		\caption{(color online) The two branches of excitations of the ideal gas: the upper branch (blue solid line), with energy given by $E = \frac{2\pi^2}{L^2} K^2$ corresponds to the elementary excitations. The lower branch  (black dashed line), with energy-momentum relation $E = \frac{2\pi^2}{L^2} K$ comes from the yrast states.
		 Momentum, as defined in the text, is dimensionless. \label{fig:two-branches}}
	\end{figure}
It is very instructive to investigate the system in the simplest case of the ideal gas.
In the case without interaction, every Fock state in the plane wave basis is already an eigenstate of the Hamiltonian \eqref{Ham}. 
The energy of the Fock state $\ket{\vec{n}} = \ket{n_{-\infty}...n_{k}...n_{\infty} }$ is
	\begin{equation}
	E(\vec{n}) =  \frac{2\pi^2}{L^2} \sum_{k= -\infty}^{\infty}\,n_k \,k^2,
	\label{eq:energy-ideal-gas}
	\end{equation}
We distinguish two characteristic types of excitations. The first ones are the {\bf elementary} excitations obtained from the ground state $\ket{n_0=N}$ by taking a single atom to momentum $K$, so the total momentum is carried by a single particle. The spectrum is given by the parabola $E = \frac{2\pi^2}{L^2} K^2$. There is also another important branch consisting of the lowest energy states at a given momentum, i.e. the yrast states. 	
To find the yrast state in this case one has to identify which set of integers $n_k$ minimizes the kinetic energy  \eqref{eq:energy-ideal-gas} but under constrained total momentum $K = \sum_{k= -\infty}^{\infty}\,k\,n_k$. One finds that the yrast state with momentum $K$ is a state with $K$ atoms occupying the plane wave with momentum $k=1$, namely the orbital  $\frac{1}{\sqrt{L}}\,e^{i 2\pi x/L}$, and the rest of them remain in the state  $\frac{1}{\sqrt{L}}$ corresponding to $k=0$:
	\begin{equation}
	\ket{{\rm yrast:}\, N, \,K} := \ket{n_0 = N-K, n_1 = K} .
	\label{eq:yrast-ideal-gas}
	\end{equation}
The spectrum of the yrast states is $E=  \frac{2\pi^2}{L^2} K$. The Eq. \eqref{eq:yrast-ideal-gas} tells us, that the yrast states are rather the {\bf collective} excitations as obtained by exciting simultaneously $K$ atoms.

These two branches, depicted in Fig. \ref{fig:two-branches}, are nothing else but the two branches of excitations found by E. Lieb \cite{Lieb1963} but in the limit $g\to 0$, both named elementary excitations in the literature. 
Apparently this nomenclature looses sense in the limit $g\to 0$, where the  type II excitations are collective.

The perturbation theory teaches us that at least for weak interaction the yrast states should be dominated by the eigenstates identified already in the non-interacting case, given in Eq. \eqref{eq:yrast-ideal-gas}, as shown in \cite{Fialko2012} and reminded in the previous section.
This is where the surprise comes --  we tried to convince the Reader, that the many-body yrast states have solitons built-in. On the other hand we see that dominant role is played by the solutions of the non-interacting case, where there is no source for the nonlinearity and henceforth no orthodox soliton can appear. How come that the solutions with nice solitonic properties, like  density notches and  phase jumps, emerge in this regime? 
Are the additional Fock states forming the yrast states, with residual weights not exceeding $0.5$\%, sufficient to build up the solitonic properties?

To answer these questions we analyze below the conditional wave function of the yrast states in the case without interaction.
We start with the statistical properties of the system in relation to a measurement.
\subsection{Multiparticle wave function vs measurement\label{sec:measurement}}

\begin{center}
	\begin{figure}[h]
\includegraphics[width=.2\textwidth]{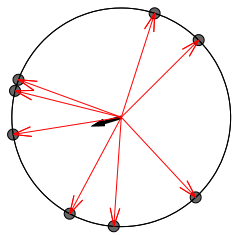}
\caption{(color online) Illustration of the definition of the center of mass (black thick arrow), being here a vectorial sum of vectors (thin red arrows) pointing to the particles. The box with the periodic  boundary condition is here interpreted as a circle.
		\label{fig:center-of-mass}}
	\end{figure}
\end{center}
In a measurement performed on the gas of $N$ atoms one obtains in fact an image of the $N$-th order correlation function \cite{bach2004}.
We reconstruct the experimental-like measurement by drawing $N$ positions from the yrast state using its probability density $|\psi (x_1,\ldots,\,x_N)|^2$ as the $N$-body distribution. To perform such drawings we use the algorithm of Metropolis.
In each "measurement" we have $N$ points, as experimentalists have on CCD cameras.  We repeat such drawing many times.
Due to the translational symmetry, the center of mass is a random variable with the rotationally uniform distribution. To reveal any hidden correlations one has to appropriately align the samples. 
We do it by rotating samples such that their centers of mass point in the same direction. 
The center of mass has to be understood here as a vector, see Fig. \ref{fig:center-of-mass}.
After such alignments we construct a histogram of particles' positions.

\begin{center}
	\begin{figure}[h]
		\includegraphics[width=0.22\textwidth]{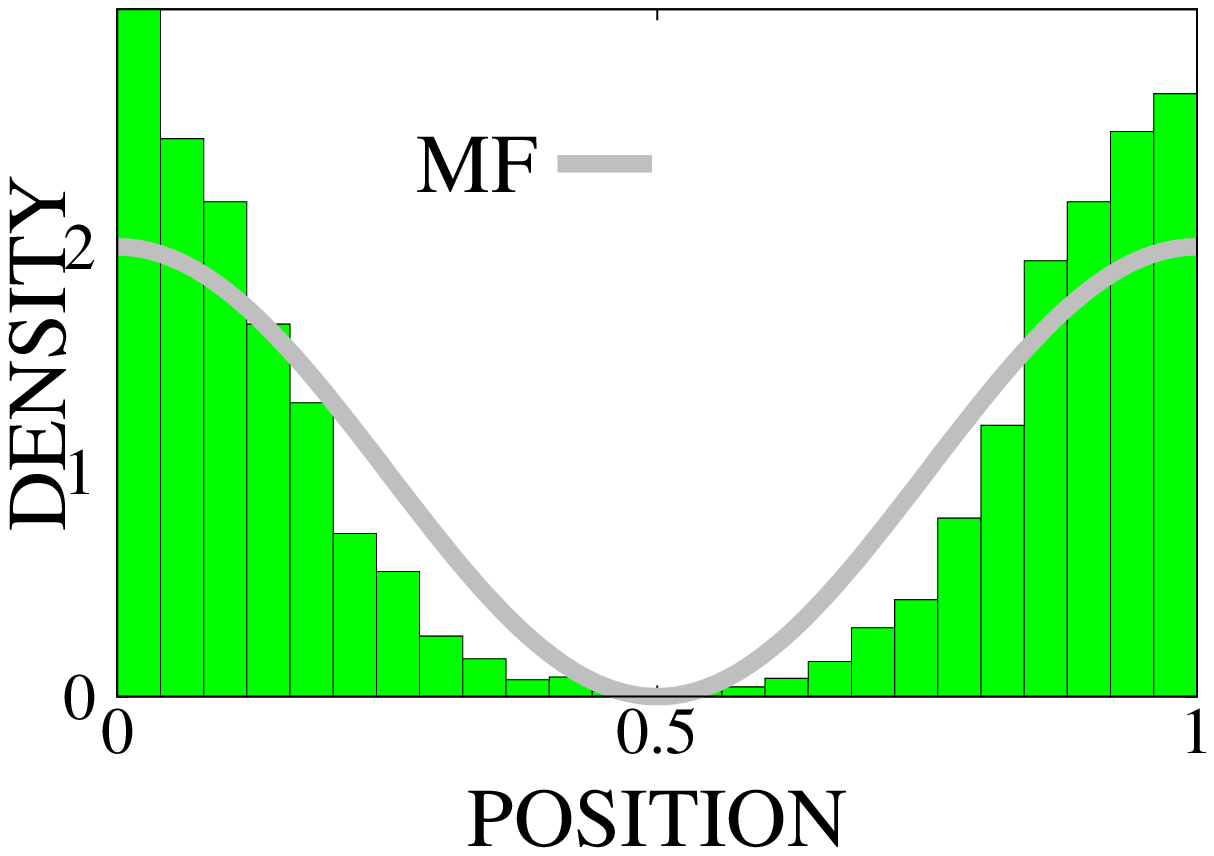}
		\includegraphics[width=0.22\textwidth]{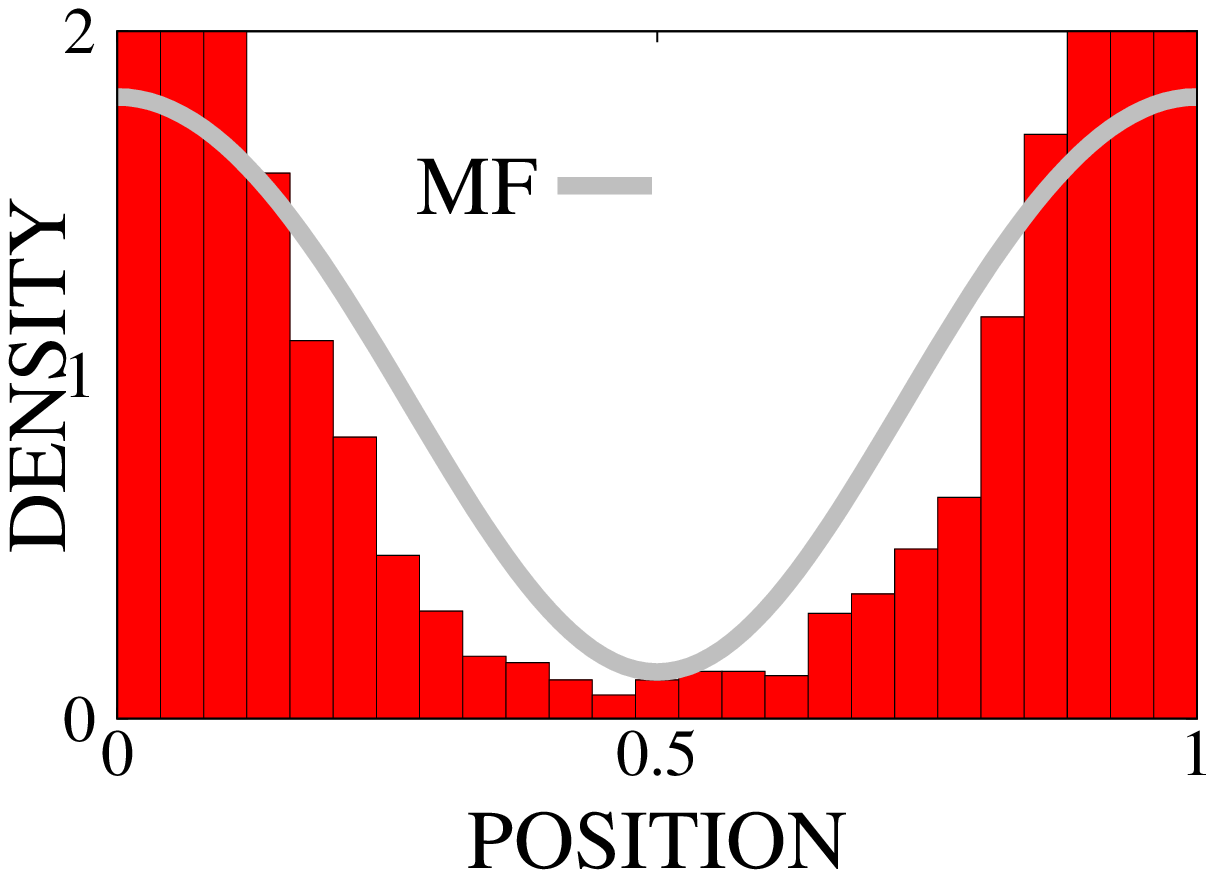}
		
		\includegraphics[width=0.22\textwidth]{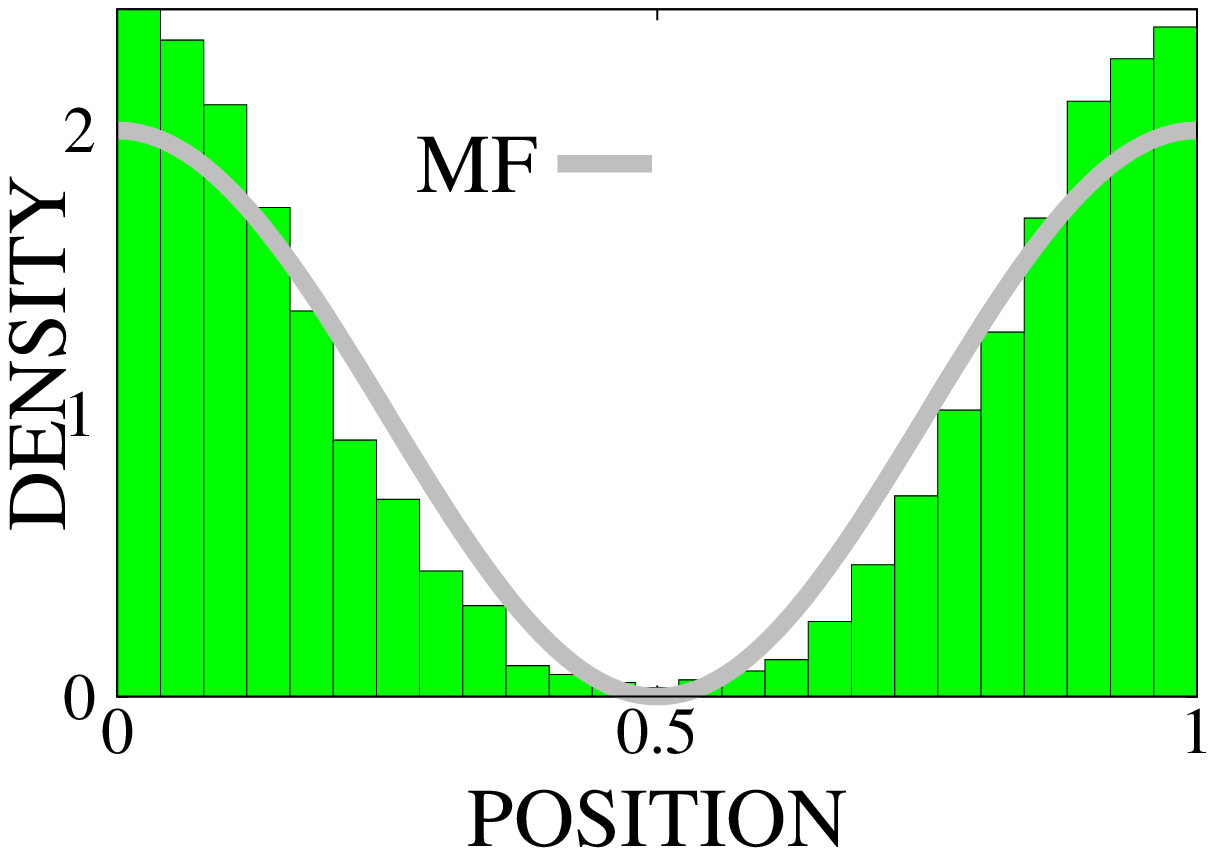}
		\includegraphics[width=0.22\textwidth]{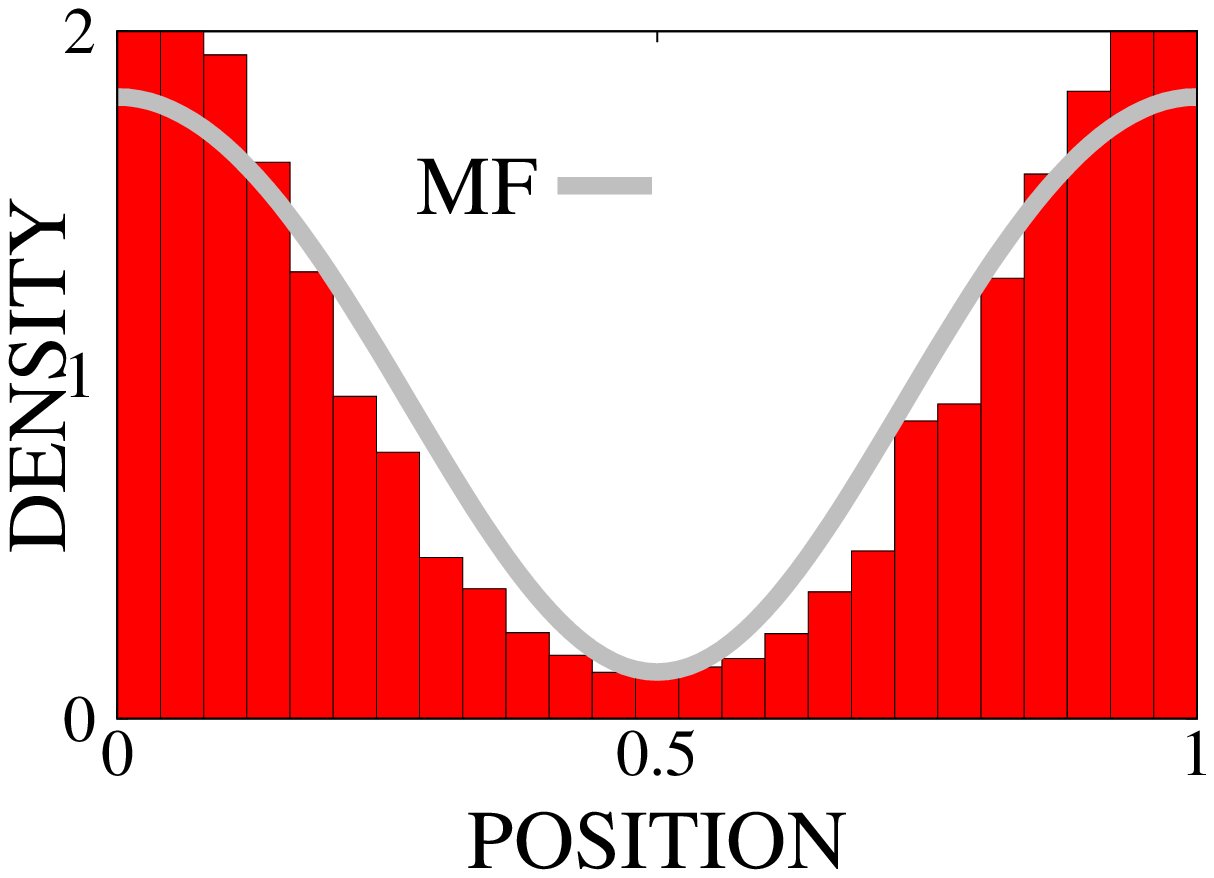}
				
		\includegraphics[width=0.22\textwidth]{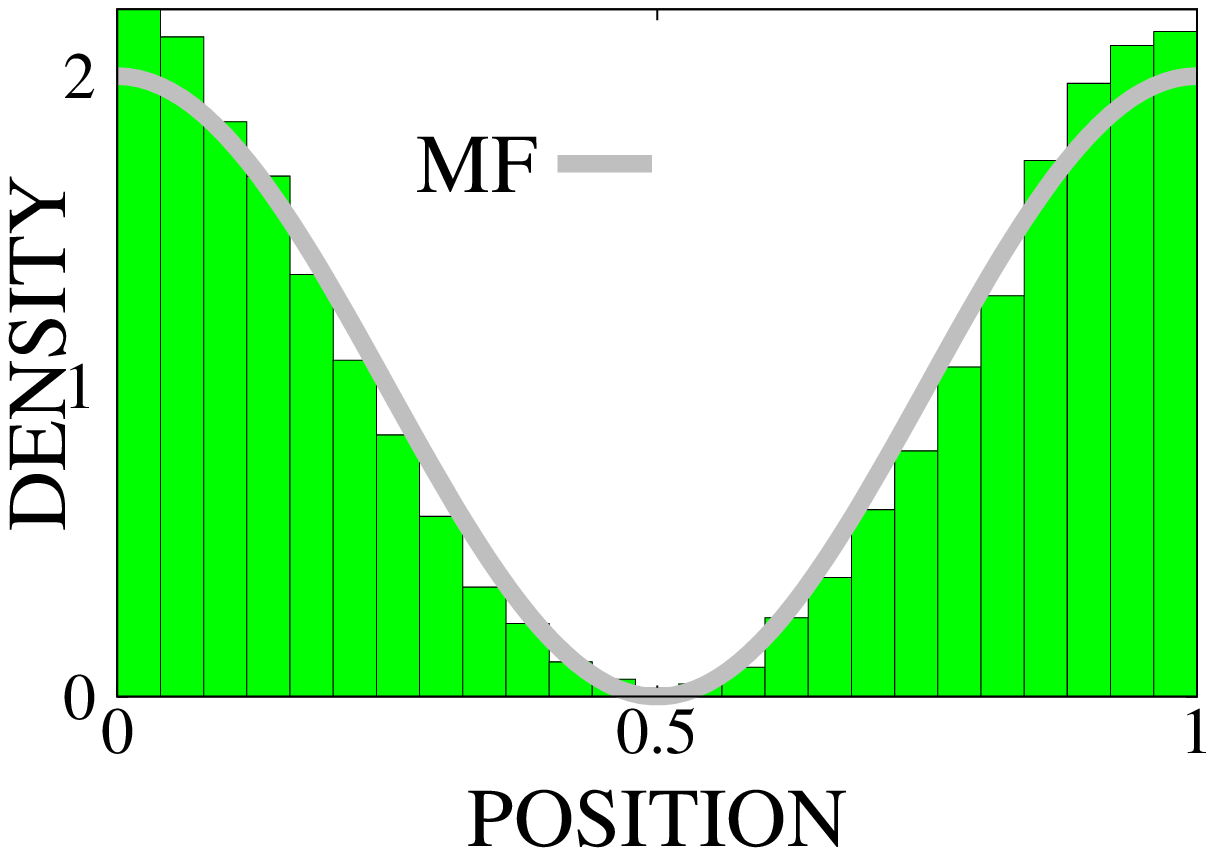}
		\includegraphics[width=0.22\textwidth]{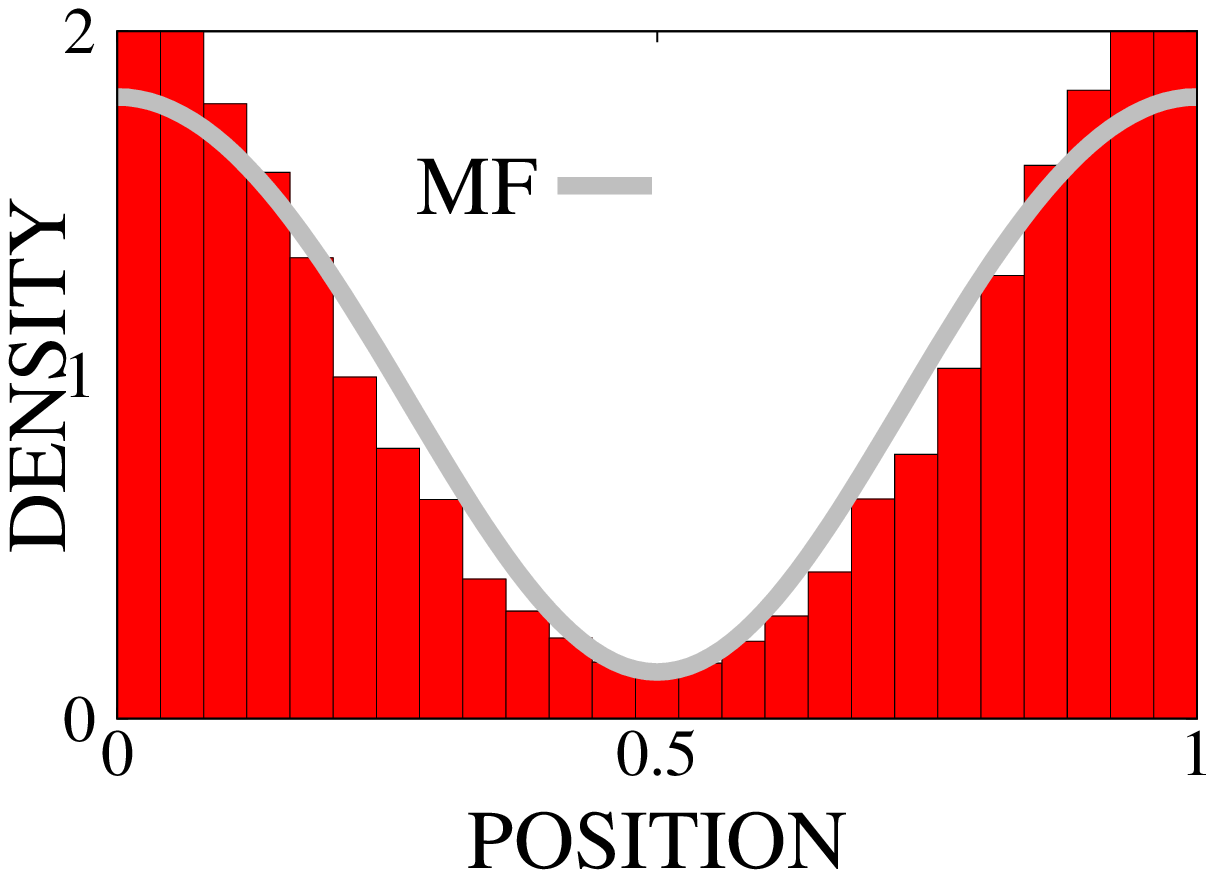}

		\includegraphics[width=0.22\textwidth]{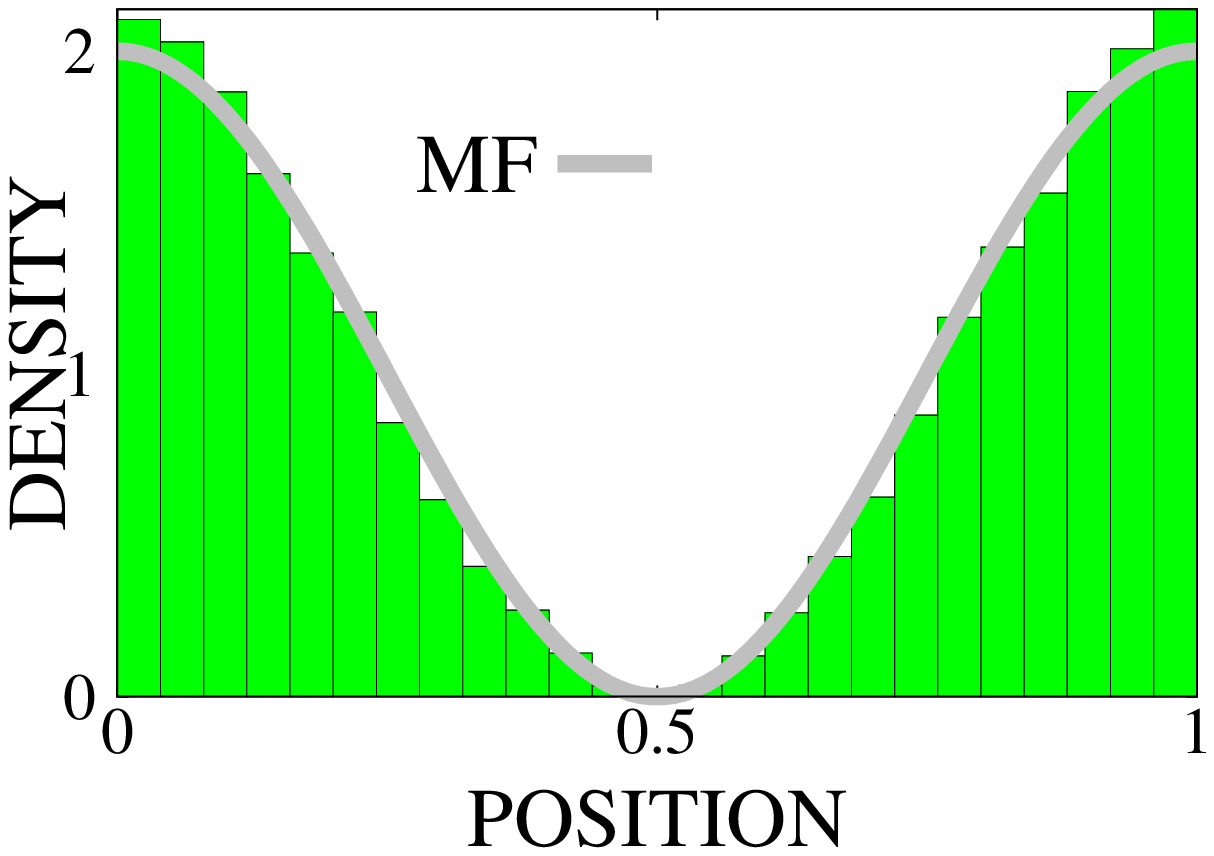}
		\includegraphics[width=0.22\textwidth]{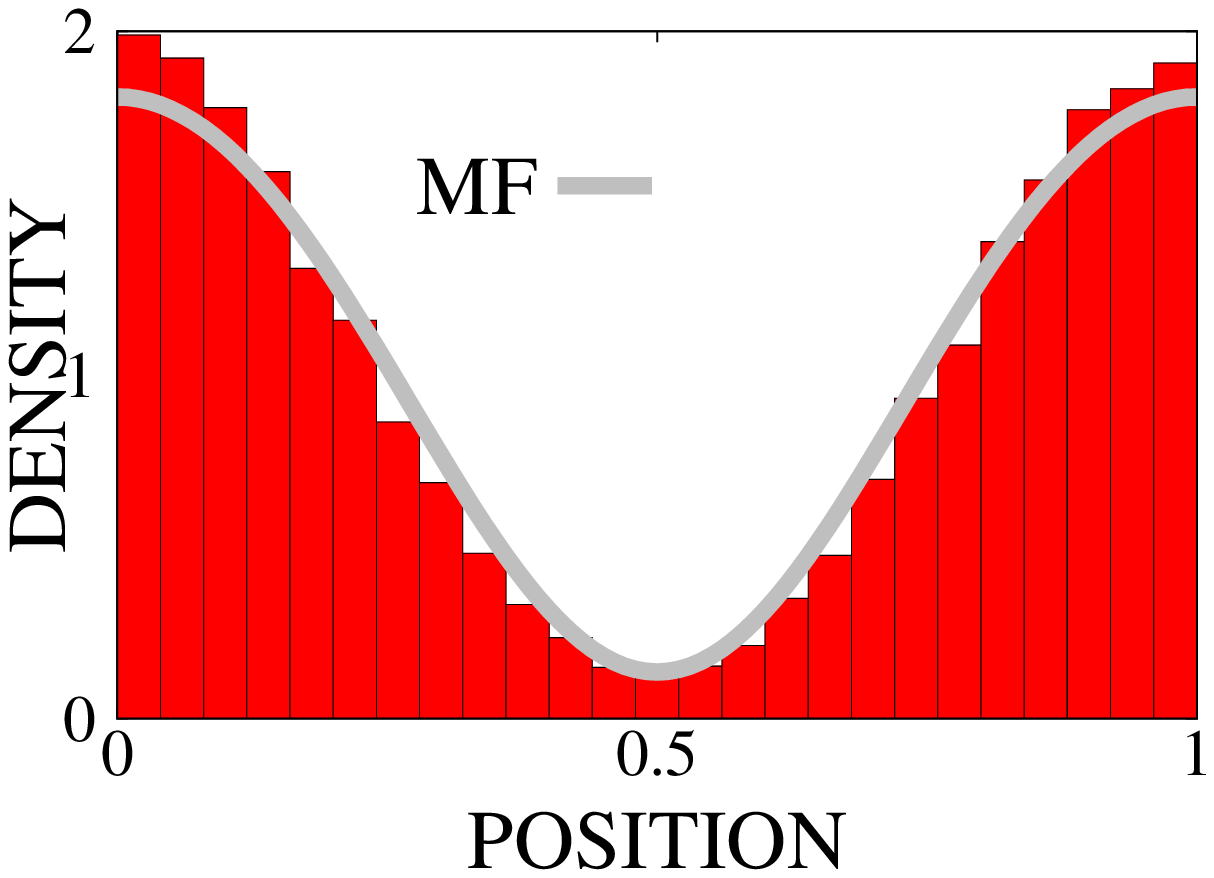}
				
				\caption{(color online) The histogram of positions drawn from the many-body yrast state of the ideal gas. The left column is for states with momentum $K=(N/2)$ corresponding to dark solitons. The right column is for the yrast state with fixed total momentum $K=N/4$. The number of atoms, from top to bottom, is $N=4$, $N=8$, $N=16$ and $N=32$. The solid gray line is the mean-field black soliton for $g=0$. In all panels the number of samples was $1000$.
				The histograms were computed after shifting particles such that their centers of mass point in he same direction  (see Fig. \ref{fig:center-of-mass}). All samples drawn from $N$-particle distribution. Position is in the unit of the box length $L$.
		\label{fig:histograms}}
	\end{figure}
\end{center}

%
%

The results are presented in Fig. \ref{fig:histograms} for yrast states of the ideal gas \eqref{eq:yrast-ideal-gas} in two cases: the  total momentum $K=N/2$ (left column with green histograms) and the total momentum $K=N/4$ (right column with red histograms). 
As $N$ grows the  positions' histograms approach the mean-field densities (in a rotated frame). 
Hence we show that even in the case of the ideal gas, one can extract from the many-body eigenstate a distribution with density notch, the same which  appears in the time dependent mean-field analysis.
Moreover the Fig. \ref{fig:histograms} demonstrates that such distribution can be extracted from the measurements. 
In the next subsection we study the black soliton-like states analytically.

\subsection{Black Solitons-like States\label{subSec:black-solitons}}
As observed in the Sec. \ref{sec:Weakly-interacting case} the many-body eigenstate minimizing the energy in the subspace with the total momentum $K = N/2$ is dominated just by the single Fock state. 
Here we focus on the conditional wave-function of this  state  to show how it leads to the density notches and  jump in the phase.
In the spatial representation this state reads:
	\begin{align}
	\psi&\bb{x_1, \, x_2,\,\ldots, x_N} := \langle x_1,\,x_2,\,\ldots x_N\ket{{\rm yrast:}\, N, \,K=N/2} \nonumber\\
	 &= \frac{1}{\sqrt{L^N\,\binom{N}{N/2}}} \sum_{\sigma } \,e^{i 2 \pi \bb{x_{\sigma(1)} + x_{\sigma(2)} + \ldots + x_{\sigma(N/2)}} /L},
	\label{eq:black-yrast-ideal-gas}
	\end{align}
where the sum is over all possible subsets of $N/2$ atoms out of $N$. 
We look at the many body wave-function conditioned to "measured" positions of $N-1$ atoms, i.e. we treat the first $N-1$ positions as parameters.
The resulting conditional wave-function  of $N$-th particle \eqref{eq:conditional-wave-function} is:
	\begin{equation}
	\psi_{\rm con}^{x_1, \, \ldots,\,x_{N-1}} (x_N) \propto S\, e^{ 2 i\pi x_N / L} + M,
	\label{eq:sol-with_M_S}
	\end{equation}
where $S$ is the sum of $\binom{N-1}{N/2-1}$ terms consisting of products of $N/2-1$ plane waves. 
Similarly the number  $M$ is the sum of $\binom{N-1}{N/2}$ terms consisting of products of $N/2$ plane waves. 
Their explicit forms, denoting the phase factors with $a_i := e^{2 i\pi x_i/L}$, are given by
\begin{eqnarray}
S &=& \sum_{\sigma \in A_{N/2-1}}  a_{\sigma (1)}\,a_{\sigma (2)} \ldots \, a_{\sigma (N/2-1)} \nonumber ,\\
M &=& \sum_{\sigma \in A_{N/2}}  a_{\sigma (1)}\,a_{\sigma (2)} \ldots \, a_{\sigma (N/2)} \label{eq:S-and-M},
\end{eqnarray}
where the sums are over all possible subsets of $N/2-1$ and $N/2$ positions from the set $\left\{x_1,\, x_2,\,\ldots, x_{N-1}\right\}$.
Note that we perform analysis for any positions of the first $N-1$ atoms, not for the ones drawn from the many-body distribution, as it was done in the previous section.
Both stochastic functions, $S$ and $M$ have the same number of terms, $\binom{N-1}{N/2} = \binom{N-1}{N/2-1} $. 
Each term from the sum in $S$ has a counterpart in $M$ due to the identity: $\bb{\prod_{i=1}^{N-1} a_i} \bb{\prod_{j=1}^{N/2 - 1} a_j}^*=\prod_{j=N/2}^{N-1} a_j$.
This leads to a conclusion that the complex number $S$ is nothing else, but the complex number  $M$ reflected and rotated on the complex plane, i.e. $\bb{\prod_{i=1}^{N-1} a_i} S^*=M$.
With this observation we can write the conditional wave-function \eqref{eq:sol-with_M_S} in a simpler form
	\begin{equation}
	\psi_{\rm con}^{x_1, \, \ldots,\,x_{N-1}} (x_N)\propto 1 + e^{2 i \pi (x_N + X) / L},
	\label{eq:twin-fock-position-explicitly}
	\end{equation}
where $X = \sum_{i=1}^{N-1}x_i - \frac{L}{\pi} {\rm Arg} \bb{M}$.
In other words we find that irrespectively of the positions of $N-1$ atoms, the yrast state \eqref{eq:black-yrast-ideal-gas} treated as a function of the $N$th atom has the form $1+e^{2 i \pi x_N / L}$ up to a shift of $x_N$ by a distance $X$ depending on all other particles.
The conditional wave-function \eqref{eq:twin-fock-position-explicitly} has a density profile $1+\cos\bb{2\pi\,x_N/L}$ mimicking the density notch known for soliton. In the position of density minimum at $X$, the phase jumps by $\pi$, again as in the black soliton known from the non-linear Schr\"odinger equation. 
These density and phase profiles coincides with the results for black soliton-like states, presented in Fig. \ref{fig:sol}.
As there is no source of non-linearity these are no real solitons  -we cannot speak about healing length or compensation between the dispersion and inter-atomic repulsion. Still we have an interesting 
conclusion: the typical properties of the soliton, density notch and the appropriate phase jump, appear already in the case without interaction. There is still a good agreement between the profiles of the conditional states and the 'solitons' found in the corresponding Schr\"odinger equation.
This agreement may seem accidental: in the naive derivation of the non-linear Schr\"odinger equation one assumes that the many-body wave function is a product state with all atoms occupying the same orbital. 
In the case of the ideal gas the conditional wave-function has indeed a form independent of all other $N-1$ positions, but up to a shift $X$. The solitonic-properties result from the Fock state $\ket{\frac{N}{2}, \frac{N}{2}}$.
This state written in the position representation,  as given in Eq. \eqref{eq:black-yrast-ideal-gas}, is not a product state. On the contrary: due to the  bosonic symmetrization each particle is correlated with all other particles, hence the state is highly correlated.
We reach a paradox: we find the mean-field solutions in the many-body state which is very far from the assumptions on which the mean-field model relies.
This paradox is strongly related to the famous debate about the interference of the Fock states \cite{Javanainen1996, Leggett1991}. The average density computed in the Fock state is uniform. On the other hand in each experimental realizations there was appearing a clear interference pattern \cite{Andrews637}, although at random position. 
It has been explained that the interference pattern arises in the course of measurements \cite{Javanainen1996, Castin1997} - the wave-function under the condition that a few particles were  measured at certain positions exhibits indeed a clear interference fringes. The appearance of the black 'solitons',  as well as the appearance of the interference fringes, can be understood within the following form of the many-body wave-function \cite{Castin2001}:
\begin{multline}
	\psi (x_1,\ldots,x_N) = \int_0^L {\rm d}X\,e^{- i\pi \bb{2\hat{K}-N} X/L} \prod_{i=1}^N \, \psi_{\rm GPE} (x_i) \label{eq:crucial}\\
	  \propto  \int_0^L {\rm d}X\,e^{- 2 i \pi  \hat{K} X/L} \bb{e^{i  \pi N X/L}} \prod_{i=1}^N\bb{1 + e^{2 i\pi x_i/L}}.
\end{multline}
In other words, the state is a superposition of the same product states but with  all possible positions, such that  the translational symmetry is preserved. 
"Measuring" a few positions would break the translational symmetry and cause 
collapse of the wavepacket onto one of the superposed states $ \prod_{i=1}^N \, \psi_{\rm GPE} (x_i+X)$, namely  
to the state \eqref{eq:twin-fock-position-explicitly}. Hence the density notch appears at a random place,  determined by the first few detected particles as discussed in \cite{Javanainen1996, Castin2001}.

We would like to mention that the Fock states we investigate are broadly discussed by the Quantum Information community. 
The state $\ket{\frac{N}{2}, \frac{N}{2}}$ is called there the twin Fock state.
It was a subject of debates if the entanglement between particles in this state can be of some importance. In some sense this entanglement is trivial, because it results from the indistinguishability of atoms and arises only due to the symmetrization.
The definite answer is due to experiments \cite{klempt2011, TFScience2017} demonstrating that the twin Fock state is useful in the interferometry, reducing the uncertaintities strongly below the "classical" shot noise limits. This was expected as the quantum Fisher Information, widely used in the context of the metrology, reaches for the twin Fock state the Heisenberg scaling $O(N^2)$ exceeding the "classical" limits by a factor $N$.

\subsection{Gray soliton-like states \label{subSec:gray-solitons}}
The conditional wave-function of the Fock state $\ket{n_0=N-K,\,n_1=K}$, i.e. a Dicke state \cite{dicke1954}, is still of the form:
	\begin{equation}
	\psi_{\rm con}^{x_1, \, \ldots,\,x_{N-1}} (x_N)\propto S_K\, e^{2 i \pi x_N / L} + M_K.
	\label{eq:sol-with_M_S-gray}
	\end{equation}
The formulas corresponding to Eq. \eqref{eq:S-and-M} read
	\begin{eqnarray}
	S_K &=& \sum_{\sigma \in A_{K-1}}  a_{\sigma (1)}\,a_{\sigma (2)} \ldots \, a_{\sigma (K-1)} \nonumber ,\\
	M_K &=& \sum_{\sigma \in A_{K}}  a_{\sigma (1)}\,a_{\sigma (2)} \ldots \, a_{\sigma (K)} \label{eq:S-and-M-gray},
	\end{eqnarray}
where $A_M$ are all subsets of $M$ positions from the $N-1$ "measured" particles.
The probability density is given by 
	\begin{eqnarray}
	|\psi^{x_1, \, \ldots,\,x_{N-1}}_{\rm con} (x_N)|^2 &=& |S_K|^2 + |M_K|^2 + \nonumber\\
	 & & + 2|S_K \, M_K |\,\cos(\phi(x_N))
	\label{eq:sol-gray-cond}
	\end{eqnarray}
where $\phi(x_N) = 2\pi \,x_N/L + {\rm Arg}\left\{S_K - M_K\right\}$. Clearly this density has to be larger than $(|S_K|-|M_K|)^2$, namely
the more differ the absolute values $|S_K|$ and $| M_K |$ the shallower is the dip in the density. We illustrate grey "solitons" in Fig. \ref{fig:histograms} (red panels) by histograms obtained for $K=N/4$, and compare them with the mean-field solution  in the non-interacting limit with the average momentum $\langle \hat{k}\rangle =1/4$. The wave-function of the mean field gray soliton converges to $1+ A\, e^{2 \pi i x}$.
\begin{center}
	\begin{figure}[h]
		\includegraphics[width=0.22\textwidth]{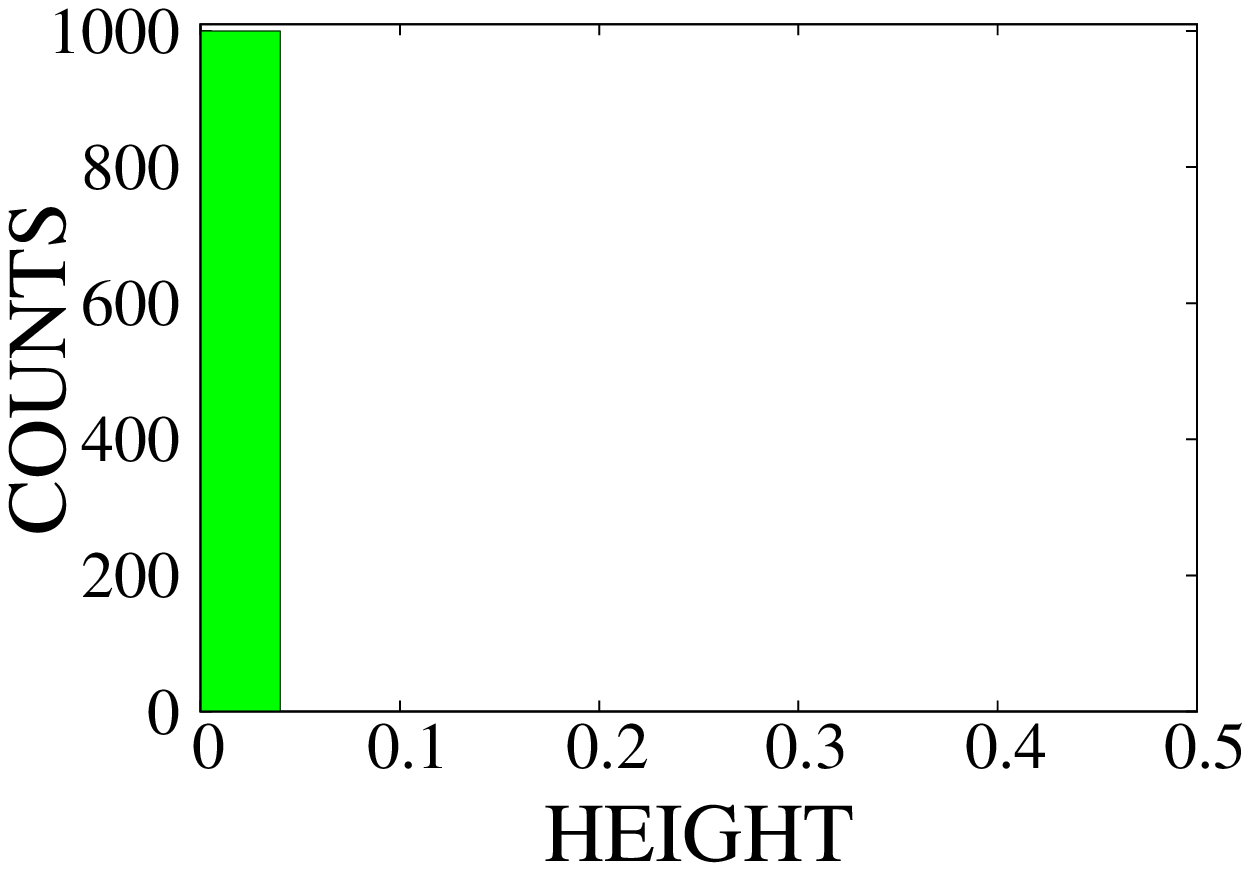}
		\includegraphics[width=0.22\textwidth]{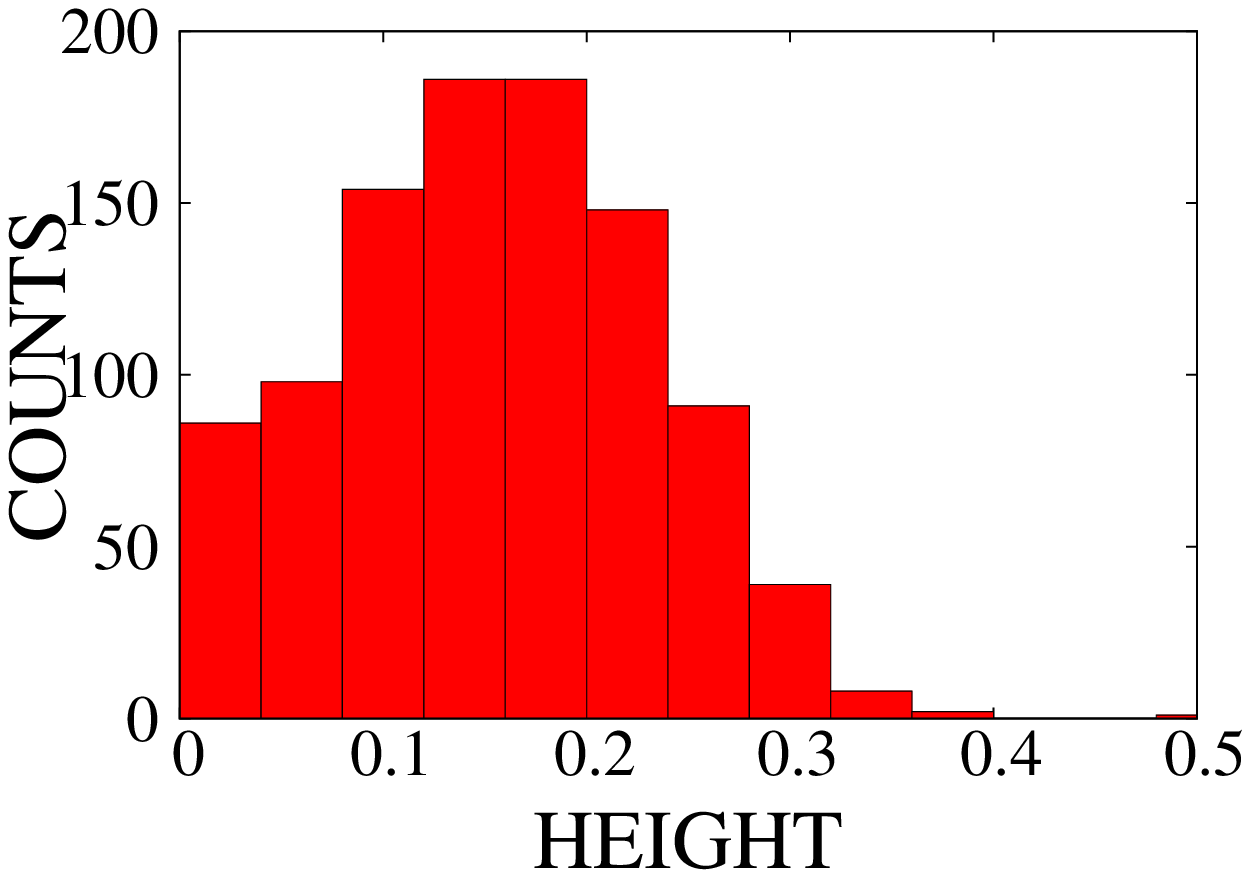}
		
		\caption{(color online) The histogram of heights of square of conditional wave function at its minimum drawn from the many-body yrast state of the ideal gas for $N=32$ atoms. The left panel: $K=N/2$ (corresponding to black soliton), the right panel $K=N/4$. In both cases the number of samples was $1000$.
			\label{fig:heights}}
	\end{figure}
\end{center}
Contrary to the black "soliton" case the form of the conditional wave-function of the gray "soliton" Eq. \eqref{eq:sol-gray-cond} is not universal. We illustrate its diversity in the right panel of Fig. \ref{fig:heights}, showing the histogram of the heights of the conditional wave-function obtained from $1000$ samples. The height equal to $0$ corresponds to the black soliton.
We note that the depth of grey solitons varies significantly from shot to shot.
\subsection{Multi-soliton-like states \label{subSec:multi-solitons}}
The superposition of two solutions of some non-linear equation usually is not the solution any more.
The situation is different for the equations supporting solitons, for which there is some sort of the superposition rule.
This is then tempting to verify if there exists multiple solitons-like solutions in the ideal gas. 
In the limit of vanishing interaction the eigenstate with two black solitons built-in  converges to the Fock state $\ket{n_{-1}=N/2,\,n_1=N/2}$. One can perform the analysis similar to the one from the previous section to obtain the conditional wave-function
	\begin{equation}
	\psi_{\rm con} (x_N) \propto \cos(2\pi (x_N - X)/L),
	\label{eq:double-soliton}
	\end{equation} 
where, as before, the shift $X$ is the random variable which depends on the positions $x_1,\ldots,\, x_{N-1}$. 
The probability density in this case is given by $\cos^2(2\pi (x_N - X)/L)$ with two local minima at positions $X+1/4L$ and $X+3/4L$.
At each node the conditional wave function \eqref{eq:double-soliton} changes sign, i.e. it has a $\pi$-jump in the phase, similarly to the black solitons.
It is easy to find the solutions with $M$ black "solitons": the many-body eigenstate with $M$ black "solitons" is  $\ket{n_{-M/2} =N/2;\, n_{M/2}= N/2}$.
\subsection{Moving "solitons" \label{subSec:moving-solitons}}
It is natural to ask if the "solitons" can move. Within the many-body picture such movement is impossible: the states we discuss are the eigenstates of the Hamiltonian, which would gain in the evolution only a global factor $e^{-i E t}$ without any physical significance. On the other hand, after breaking the symmetry by fixing the first $N-1$ positions we obtained a conditional wave-function which  is not a stationary solution of the mean field model. 
The equivalent of a single black soliton is $\psi_{\rm con} (x_N) \propto 1 + e^{i 2 \pi (x_N - X) /L}$, namely it is a superposition of two plane waves with the energies $E_0 = 0$ and $E_1 = \frac{2 \pi^2}{L^2}$. Hence the state evolves in time
\begin{eqnarray}
\psi_{\rm con} (x_N, t) &\propto& 1 + e^{i 2 \pi (x_N - X)/L - i E_1 t} \nonumber\\
 &=& 1 + e^{i 2 \pi (x_N  - v t - X ) /L},
\end{eqnarray}
with the velocity $v=\pi/L$, as the black soliton in the case of periodic boundary conditions should move. 
Similar analysis for two black solitons shows that they are not moving at all, again exactly like in the mean field picture.

\begin{figure}[h]
		\includegraphics[width=0.22\textwidth]{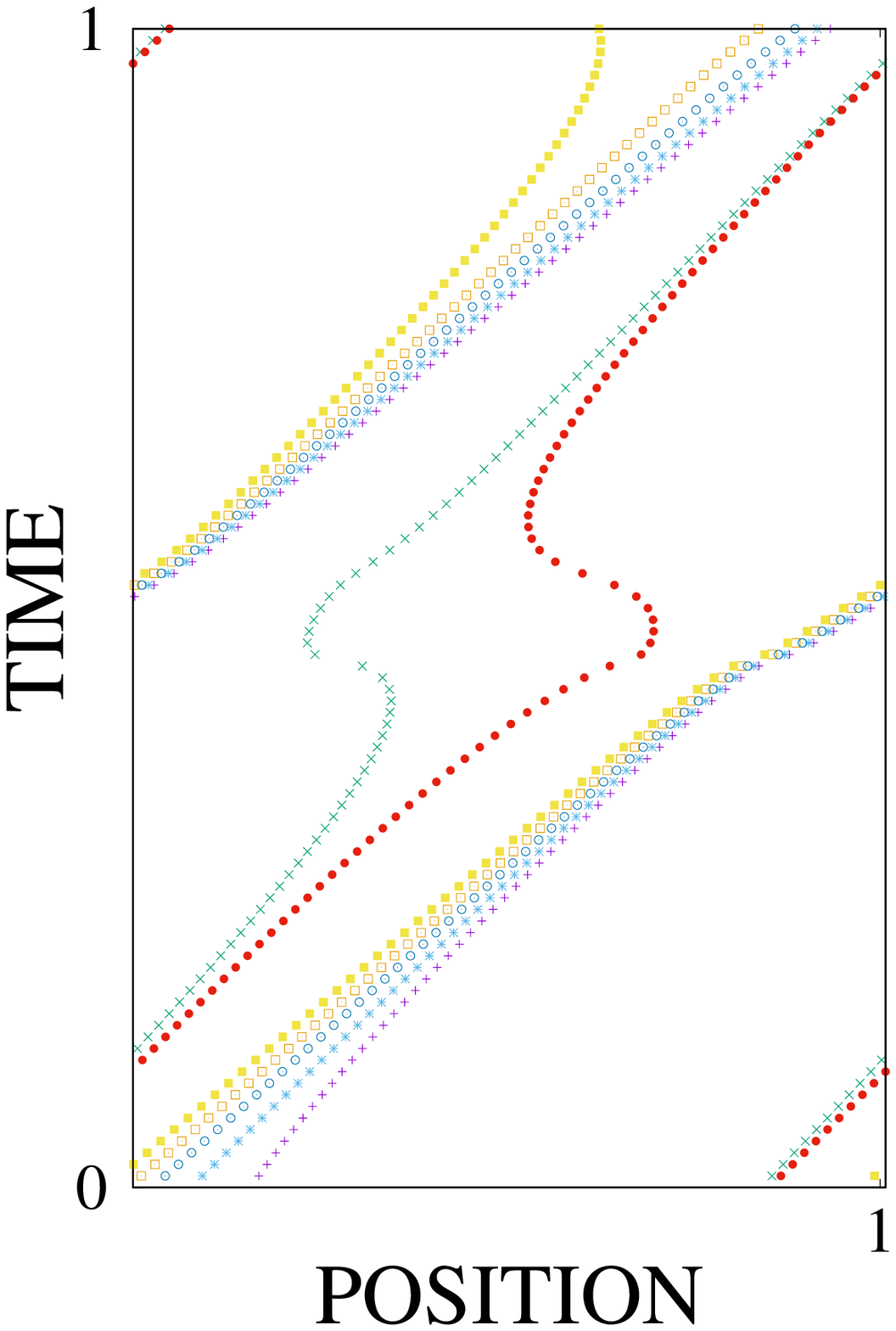}
		\includegraphics[width=0.22\textwidth]{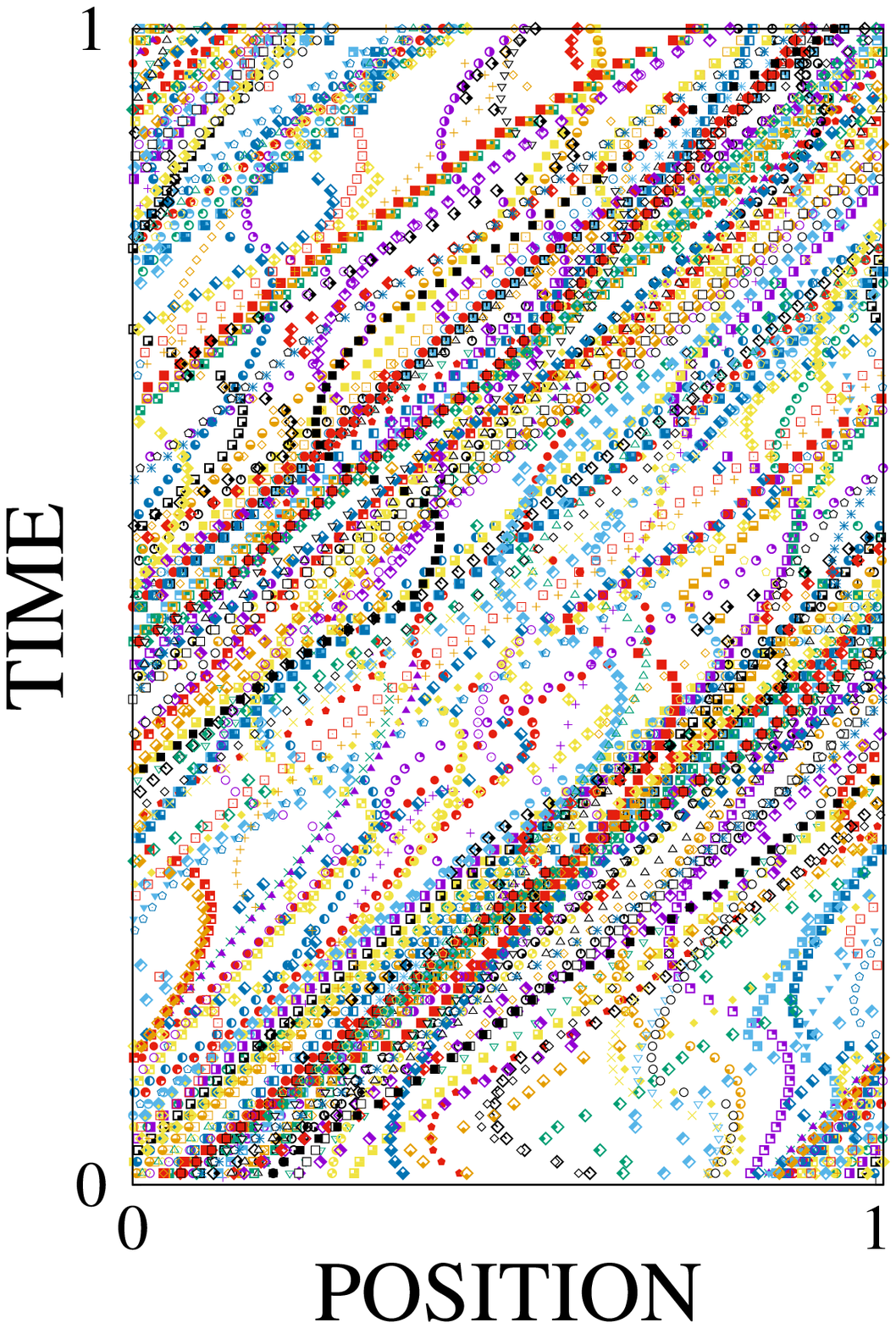}

		\includegraphics[width=0.4\textwidth]{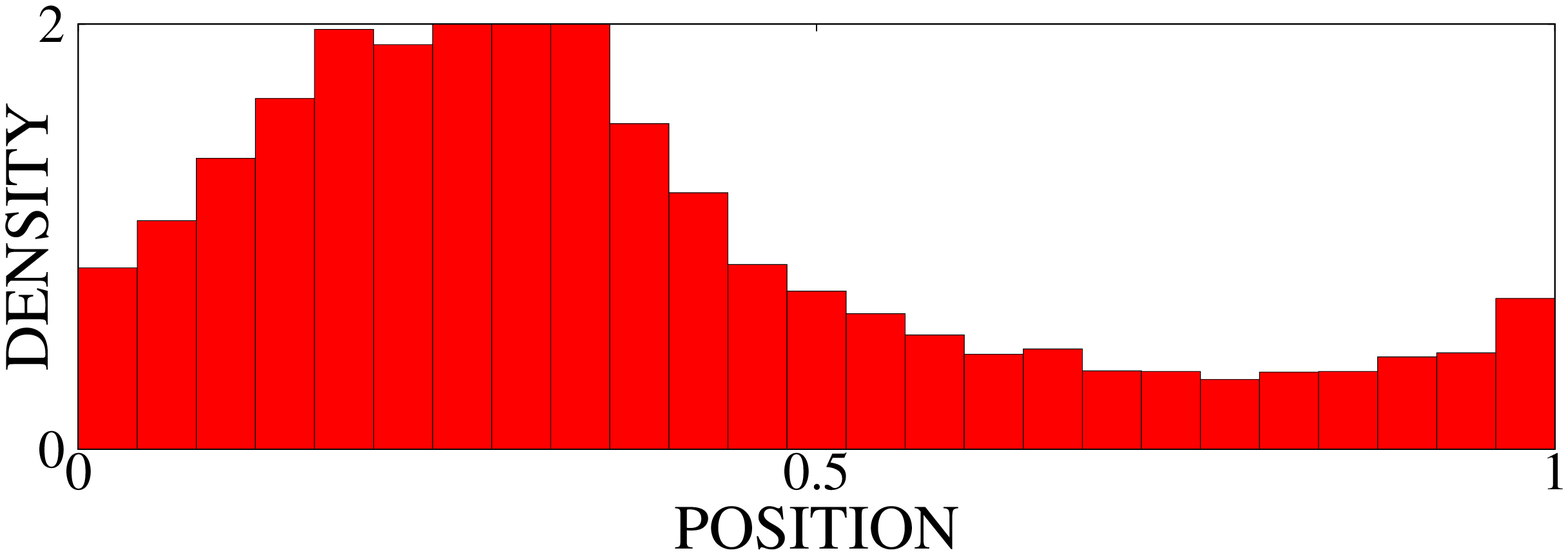}
		\includegraphics[width=0.4\textwidth]{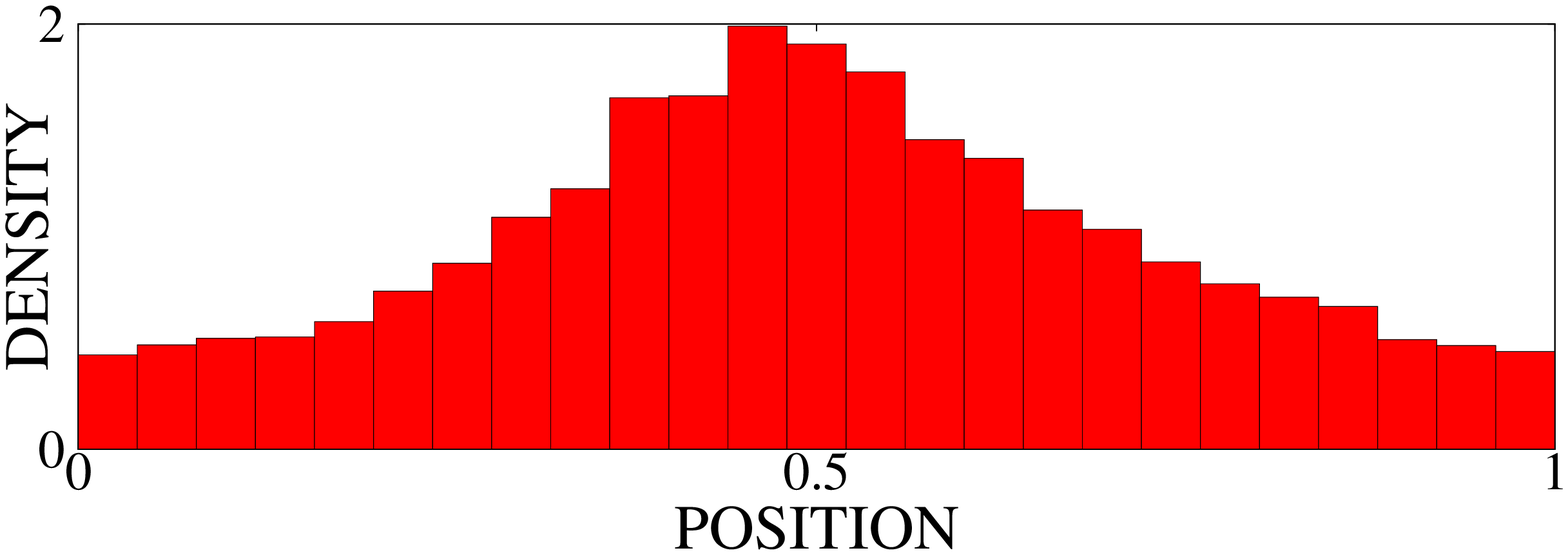}		
		\includegraphics[width=0.4\textwidth]{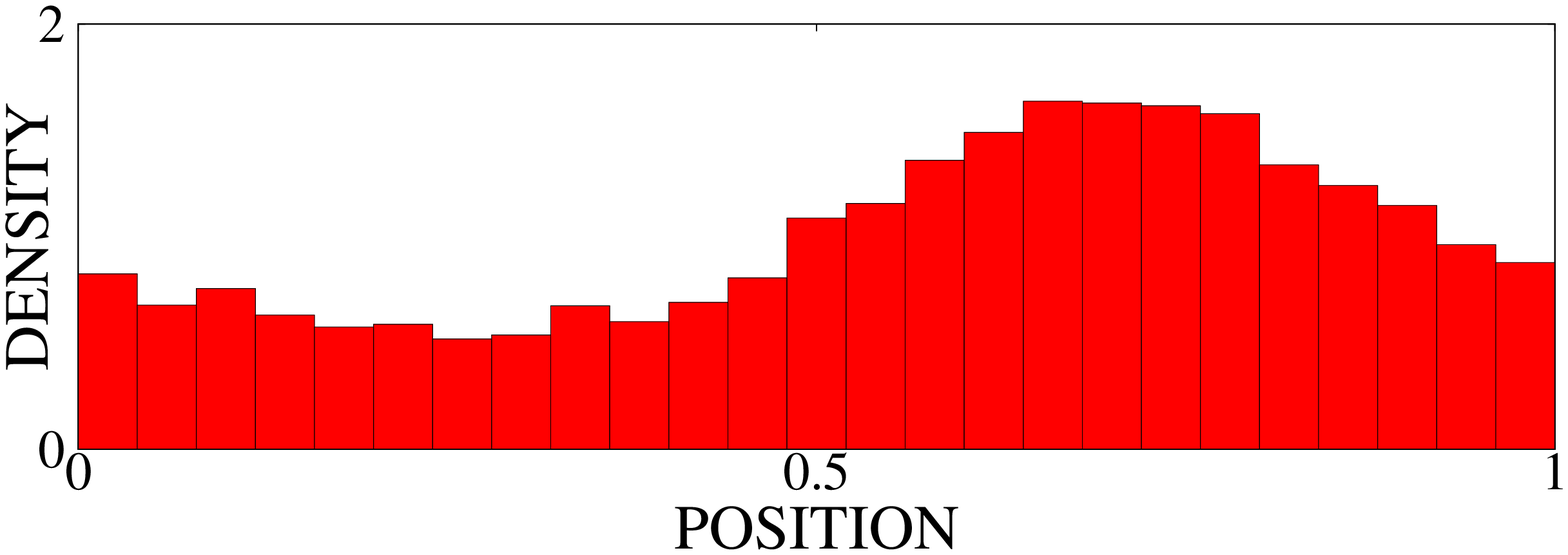}
		\includegraphics[width=0.4\textwidth]{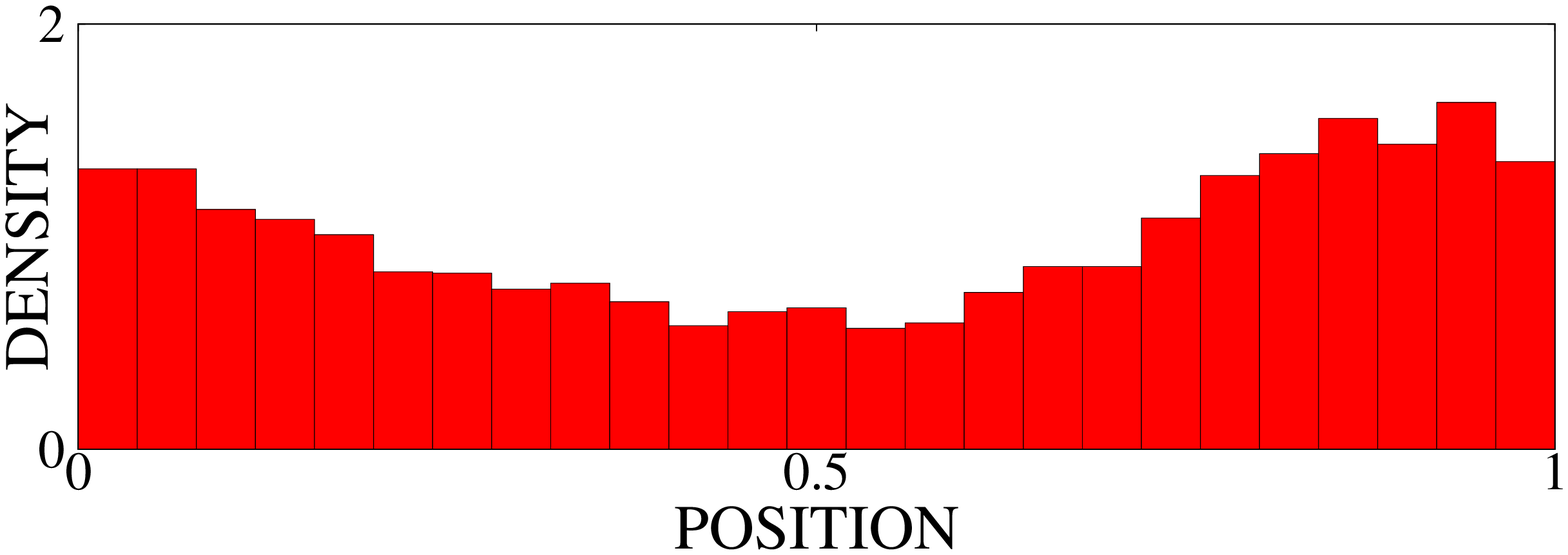}		
					
		\caption{(color online) Bohmian trajectories. Top left: Trajectories of all $N=8$ Bohmian particles from a single sample. Top right: $80$ trajectories obtained from $10$ samples. The initial positions were shifted to match the density notches, as in Fig. \ref{fig:histograms}. The histograms were obtained from positions of the Bohmian particles obtained in $1000$ realizations at times, from top to bottom $T = 0.1,\,0.2,\,0.3,\,0.4$. Position is in the unit of the box length $L$. Time is dimensionless, as in Eq \eqref{eq:non-linear-Schrodinger-equation}.
		\label{fig:bohmian}}
\end{figure}

Naively, to obtain a motion of solitons one would just evolve in time the corresponding conditional wave-functions. This, however, fails completely in case of the gray 'solitons', which within such procedure would move with the speed $\pi/L$, as the black solitons.
 To see the solitonic motion we use more sophisticated method: the Bohmian interpretation of the Quantum Mechanics \cite{Bohm1952, Bohm1952Second, Oriols2014}.
In the Bohmian picture one represents the state as a collection of $N$ point-like masses moving with the time dependent velocities.
Their initial positions should be drawn from the $N$-body probability distribution $|\psi(x_1,\ldots, x_N)|^2$.
They move similarly to the Newtonian particles, but with the velocities depending on all other particles. The velocity of the $l$th particle is given by
	\begin{equation}
	v_l =  \,{\rm Im} \left\{ \,\frac{\partial_l \psi(x_1,\ldots, x_N)}{\psi (x_1,\ldots, x_N)}\right\},
	\end{equation}
where $\partial_l := \frac{\partial}{\partial \, x_l}$ is the partial derivative with respect to the $l$-th particle.

In the case of the gray soliton-like state \eqref{eq:sol-with_M_S}, the velocity of the $N$-th particle reads:
\begin{equation}
v_N = -\frac{\pi}{L} + \frac{\pi}{L} \frac{|M_K|^2-|S_K|^2}{|\psi^{x_1, \, \ldots,\,x_{N-1}}_{\rm con} (x_N)|^2},
\end{equation}
where the probability density appearing in the denominator is given in Eq. \eqref{eq:sol-gray-cond}. The bohmian particle moves then with the velocity depending  on the local density (accelerating under the density-notch) and the total solitonic depth encoded in the parameters $S_K$ and $M_K$.

The results are presented in Fig. \ref{fig:bohmian} in the case of $N=8$ atoms and the initial state $\ket{n_0=5, n_1=3}$. We show there the examples of trajectories of all $8$ particles, drawing it once (the left panel) and then $10$ times (the right panel). At few chosen instants of time we reconstructed the histograms, using $1000$ samples of the initial positions. We observe that the solitons are moving from left to right, but also the corresponding density notch in the  histogram smears for longer evolution time. This can be understood already from Fig. \ref{fig:heights} which
shows that the states $\ket{n_0=3N/4, n_1=N/4}$ are rather collections of states with different velocities, what cause a dispersion shown in Fig. \ref{fig:bohmian}.

\section{Validity Range\label{sec:validity-range}}
\begin{figure}
	  \includegraphics[width=0.48\textwidth]{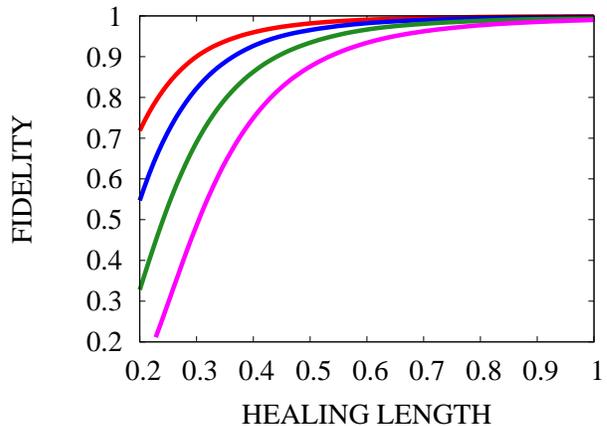}
	  \caption{(color online) Fidelity between the yrast state with momentum $K=N/2$ obtained numerically from Eq. \eqref{Ham} and the twin-Fock state $\ket{N/2, N/2}$ as a function of the healing length $1/\sqrt{g N/L}$. Number of atoms, from top to bottom $N=8,\, 16,\,32,\,64$. Position is in the unit of the box length $L$.
	  \label{fig:validity-range}}
\end{figure}
Finally we ask the question: how long the non-interacting yrast states approximate well the eigenstates of the interacting system.
The many-body yrast states of the Lieb-Liniger model are constructed from the plane waves with $N$ pairwise different quasi-momenta. 
These quasi momenta are solutions of the set of transcendental Bethe equations.
On the other hand we know already that for the ideal gas the exact solution is the single Fock state, with $N-K$ atoms in momentum $0$ and $K$ atoms in momentum $k=1$. Then one can ask the question what are the Lieb's quasi-momenta in the limit of vanishing interaction.
We checked, in the case of the black soliton, that half of the quasi-momenta known from the Lieb solutions converge to $0$ and the second half to $1$. It means, regarding the previous sections, that the quasi-momenta become the true particle momenta. 
The quasi-momenta are not analytic functions of the interaction strength at $g=0$, they converge with the rate $\sqrt{g}$. 
Since the number of equations for quasimomenta grows with $N$, the small parameter should be rather  the inverse of the healing length $\sqrt{gN/L} = 1/\xi$.
We verify this predictions in  Fig. \ref{fig:validity-range}, where we show the fidelity between the yrast state with momentum $K=N/2$ and the twin-Fock state $\ket{N/2, N/2}$. 
Note that the ideal gas approximation is fairly accurate even for the healing length significantly shorter than the size of the box.
This agrees with the analysis in \cite{Fialko2012}.

We stress that there may be still a  correspondence between the conditional wave-functions and the mean-field solitons even for the healing lengths much shorter than $L$ \cite{syrwid2015}. Only the solitonic features can not be explained within the ideal gas picture used in the previous section.

\section{Conclusions}
Since the seminal work of E. Lieb \cite{Lieb1963} there are numerous studies of the relation between the two descriptions of $N$ interacting bosons on a ring: the nonlinear mean-field model and the more fundamental linear many-body description. 
We investigate the type II Lieb's elementary excitations in the limit of vanishing interaction strength. These excitations converge simply to the Fock states in the plane-wave  basis, in particular the many-body black soliton becomes the twin Fock state. 
In the Fock states the particles are strongly correlated, but only due to the bosonic statistics.
As in the paper \cite{syrwid2015} we start with the $N$-body eigenstates, from which we obtain a single-body wave-functions conditioned to the first $N-1$ particles measured.
We find that the conditional wave-function has the typical solitonic properties - the density notches with appropriate phase jumps.
Of course the soliton-like states are not the true solitons - there is no nonlinearity which would dictate the width of the objects. Adding interaction such that the healing length decreases significantly below the size of the box would just shrink the  density notches.

Our findings open at least two avenues to study. As the twin-Fock states are already produced experimentally, one can ask if it is possible to transform them into soliton-like states. The interesting problem is the correspondence between the states created via phase imprinting on the Bose-Einstein condensate and the real many-body solitons. At least for ideal gas it is clear that such experimental procedure does not lead to the yrast states -- the phase imprinting would keep the multiparticle wave-function in the product state, whereas the yrast state is the highly entangled twin-Fock state.

We acknowledge fruitful discussions with A. Syrwid, A. Sinatra and Y. Castin. This work was supported by the (Polish) National Science Center Grants 2016/21/N/ST2/03432 (R.O.), 2014/13/D/ST2/01883 (K.P.)  and 2015/19/B/ST2/02820 (K.R. and W.G.).


%

\end{document}